\documentclass[a4paper,11pt]{article}

\usepackage{jheppub_nt} 

\usepackage{xcolor}
\usepackage{bm}
\usepackage{array}
\usepackage{multirow}
\usepackage{extarrows}
\usepackage{graphicx}
\usepackage{setspace}
\providecommand{\tabularnewline}{\\}
\usepackage{babel}

\title{Parity-violating scalar-tensor theory and the Qi-Xiu}

\author[a,b]{Yu-Min Hu}
\author[c,1]{and Xian Gao \note{Corresponding author.}}

\affiliation[a]{Deep Space Exploration Laboratory/School of  Physical Sciences, 
University of Science and Technology of China, Hefei, Anhui 230026, China}
\affiliation[b]{CAS Key Laboratory for Researches in Galaxies  and 
Cosmology/Department of Astronomy, School of Astronomy and Space Science, 
University of Science and Technology of China, Hefei, Anhui 230026, China}
\affiliation[c]{School of Physics and Astronomy, Sun Yat-sen University, Zhuhai 519082, China}

\emailAdd{yumin28@ustc.edu.cn}
\emailAdd{gaoxian@mail.sysu.edu.cn}

\abstract{We investigate the parity-violating scalar-tensor theory and pay special attention to terms that are free of the Ostrogradsky ghost in the unitary gauge, i.e., when the scalar field possesses a timelike gradient. We exhaustively identify the generally covariant scalar-tensor theory (GST) monomials with parity violation up to $d=4$, where $d$ is the total number of derivatives in the unitary gauge. According to the correspondence between GST terms and the spatially covariant gravity (SCG) terms in the unitary gauge, we also exhaustively identify the SCG monomials with parity violation up to $d=4$, where the Lie derivatives of the extrinsic curvature and the lapse function are necessarily introduced. We find a total of 9 independent parity-violating SCG monomials, of which 7 contain no higher-order Lie derivatives and are thus automatically free of ghosts, while 2 involve Lie derivatives of the extrinsic curvature and the lapse function and are thus potentially dangerous. By explicitly deriving their generally covariant correspondence, we obtain 7 independent scalar-tensor terms dubbed the ``Qi-Xiu'' Lagrangians, which are the most general parity-violating scalar-tensor theories that are ghost-free in the unitary gauge up to $d=4$. Our results include the existing theories in the literature, such as the Chern-Simons term and the chiral scalar-tensor theories, as special cases.}

\begin{document} 
\maketitle

\section{Introduction}

While the dark sector of current cosmology remains insufficiently understood, new physics beyond general relativity (GR) and the standard model of particle physics, including the strong, electromagnetic, and weak interactions, is required.
Various observable effects, such as helicity asymmetry and beta decay, have been observed, indicating the presence of parity violation in weak interactions \cite{Lee:1956qn,Wu:1957my}.
In the era of precision cosmology, the polarization features of the cosmic microwave background reveal clues about parity violation \cite{Minami:2020odp,Eskilt:2022cff,Komatsu:2022nvu,Caloni:2022kwp,Nilsson:2023sxz} beyond the standard model.
Consequently, when turning our attention to gravity, it is natural to question if there are any parity-violating effects in gravitational interactions.

From a theoretical perspective, incorporating parity-violating terms opens up new avenues for understanding fundamental aspects of gravity.
At the quantum level, parity-violating terms would contribute to the renormalization and ultraviolet behavior of the theory, thus potentially providing insights into the quantization of gravity or the final unified theory \cite{Hojman:1980kv,Witten:1984dg,Green:1984sg,Mavromatos:2004sz,Freidel:2005sn}.
From an observational perspective, the detection of gravitational wave events \cite{LIGOScientific:2016aoc,LIGOScientific:2017vwq}, together with forthcoming gravitational wave experiments \cite{LISA:2017pwj,Kawamura:2011zz,Sathyaprakash:2012jk,Hu:2017mde,Ruan:2018tsw,TianQin:2015yph,Li:2017drr}, opens a new window and provides an opportunity to directly test these parity-violating theories and models.
Different polarization modes in a cosmological background would be exhibited in these parity-violating theories, and modifications to the dispersion relations would also be introduced \cite{Alexander:2017jmt,Zhao:2019xmm}.
Moreover, as corrections to standard inflationary models, parity-violating terms change the polarization modes of the resulting primordial gravitational waves \cite{Satoh:2007gn,Satoh:2010ep,Creminelli:2014wna,Cannone:2015rra}, which should produce non-vanishing TB/EB correlations in the cosmic microwave background polarization \cite{Lue:1998mq,Saito:2007kt,Contaldi:2008yz,Gluscevic:2010vv,Sorbo:2011rz}.
The resulting behavior of GWs in different parity-violating gravity models has been extensively studied \cite{Masui:2017fzw,Bartolo:2018elp,Nishizawa:2018srh,Mylova:2019jrj,Qiao:2019hkz,Qiao:2019wsh,Orlando:2020oko,Chen:2022wtz,Qiao:2022mln,Zhang:2022xmm,Omiya:2023rhj,Feng:2023veu,Zhang:2023scq,Li:2024fxy}.

For the purpose of constructing gravitational theories with parity violation, the parity-violating terms can enter the gravitational action through various approaches.
A certain class of quintessence or axion fields \cite{Carroll:1989vb,Pospelov:2008gg,Finelli:2008jv} would generate such a parity-violating feature by coupling the pseudo-scalar with electromagnetism.
Another notable example is the four-dimensional Chern-Simons modified gravity \cite{Chern:1974ft,Deser:1981wh,Jackiw:2003pm,Alexander:2009tp}, in which the Pontryagin term with a coefficient depending on a scalar field is added to the standard Einstein-Hilbert action.
Generalization of Chern-Simons gravity in the more general scalar-tensor theories is considered in \cite{Crisostomi:2017ugk}, in which various couplings between the Riemann tensor and covariant derivatives of the scalar field up to the second order are introduced.
A totally antisymmetric Levi-Civita tensor (volume form) $\varepsilon_{\mu\nu\rho\sigma}$ appears in the action and is used to construct various parity-violating terms.
One lesson from the construction of these parity-violating terms in the framework of scalar-tensor theory is that the resulting theories are ghost-free only if the scalar field possesses a timelike gradient so that the so-called unitary gauge can be taken.

On the other hand, generally covariant scalar-tensor (GST) theories in the unitary gauge take the form of metric theories with Lorentz violation and, more specifically, spatially covariant gravity (SCG) theories respecting only spatial covariance. The SCG theories have simpler forms and are, in fact, more convenient in the cosmological context.
Well-studied examples, although originally motivated by different purposes, include the Ho\v{r}ava-Lifshitz gravity \cite{Horava:2008ih,Horava:2009uw,Sotiriou:2009gy,Sotiriou:2009bx,Germani:2009yt,Blas:2009qj,Wang:2017brl,Chagoya:2018yna,Barausse:2021dza}, as well as the effective field theory of inflation/dark energy \cite{Creminelli:2006xe,Cheung:2007st,Creminelli:2008wc,Gubitosi:2012hu,Bloomfield:2012ff,Gleyzes:2014rba}. The parity-violating effects in Ho\v{r}ava gravity have been investigated in \cite{Pereira-Dias:2011rva,Wang:2012fi,Gong:2021jgg}.
A general class of SCG theories was proposed in \cite{Gao:2014soa,Gao:2014fra} as an alternative and more unifying approach to introducing a scalar degree of freedom and generalizing the scalar-tensor theories.
It was further extended by introducing a dynamical lapse function \cite{Gao:2018znj,Gao:2019lpz,Lin:2020nro}, nonmetricity \cite{Yu:2024drx}, as well as an auxiliary scalar field \cite{Gao:2018izs,Wang:2024hfd}.
Constraints from cosmological perturbations and gravitational waves on SCG have been explored in \cite{Fujita:2015ymn,Gao:2019liu,Zhu:2022dfq,Zhu:2022uoq,Zhu:2023wci}.

There are several advantages to working in the framework of SCG theories.
Firstly, in the case of a single scalar degree of freedom, there is a one-to-one correspondence between theories of GST and SCG through gauge-fixing/gauge-recovering procedures.
Based on the classification of GST monomials in \cite{Gao:2020juc}, such a correspondence has been discussed in detail in \cite{Gao:2020yzr,Hu:2021bbo}, and the explicit linear mappings between GST and SCG monomials are developed in \cite{Gao:2020qxy} (see also \cite{Joshi:2021azw,Joshi:2023otx}).
In other words, SCG should not be viewed merely as Lorentz-violating theories but as a general and unifying description of scalar-tensor theories.
Secondly, thanks to its dependence only on spatial diffeomorphism, the separation of temporal and spatial derivatives allows us to generally include the parity-violating terms with odd-order spatial derivatives and linear terms of higher-order time derivatives in the action without introducing ghost DoFs.
This is also reflected in the fact that the absence of ghosts in SCG (i.e., in the unitary gauge) is a necessary condition for the corresponding GST theory to be ghost-free (i.e., in any gauge) \cite{Hu:2021bbo}.
The specific behavior of theories that are degenerate in the unitary gauge is discussed in \cite{DeFelice:2018ewo,DeFelice:2021hps}.
Therefore, SCG provides a unifying framework to study gravity theories with parity violation in a systematic manner.

In this work, we aim to investigate parity-violating scalar-tensor theories without ghosts in the unitary gauge.
We concentrate on the GST Lagrangians of the polynomial type.
Instead of starting from the most general GST polynomial and studying its behavior in the unitary gauge so that all the ghost-like terms get canceled, we will make use of the correspondence between GST and SCG.
That is, we will construct the most general SCG monomials with parity violation and then find their generally covariant correspondence.
The parity-violating SCG monomials have been studied in \cite{Gao:2020yzr} without Lie derivatives of the lapse function and the extrinsic curvature.
However, these terms naturally arise in GST terms in the unitary gauge.
Therefore, a complete analysis of both GST and SCG monomials up to $d=4$ with $d$ being the number of derivatives in the unitary gauge is needed.
Then, by making use of the covariant correspondence, we can get the desired parity-violating scalar-tensor theories without ghosts in the unitary gauge.
This work is thus devoted to these issues.

The paper is organized as follows. In Sec. \ref{sec:scg}, we briefly review the general framework of SCG theory. We also show a specific SCG model in Sec. \ref{subsec:ndlapse} with parity-violating terms of the polynomial type as an illustration of the construction and classification of the SCG monomials.
In Sec. \ref{sec:pvscg}, we list the basic building blocks for SCG by introducing new operators involving Lie derivatives of the lapse function and the extrinsic curvature. We then exhaust all the possible parity-violating SCG monomials up to $d=4$ with $d$ being the total number of derivatives.
In Sec. \ref{sec:gfgst}, we connect the ghost-free SCG model with parity-violating scalar-tensor theories. In particular, we get a set of 7 parity-violating scalar-tensor Lagrangians that are ghost-free in the unitary gauge.
Finally, Sec. \ref{sec:con} is devoted to conclusions.

Throughout this paper, we use $i,j,k,\cdots$ to denote spatial indices in a coordinate basis, and $a,b,c,\cdots$ to denote spacetime indices in a general basis.
Curvature tensors such as $R_{abcd}$ and $R_{ab}$ denote the 4-dimensional quantities, while ${}^{3}\!R_{ab}$ and ${}^{3}\!R_{ij}$ etc. denote the 3-dimensional (spatial) quantities.

\section{Spatially covariant gravity} \label{sec:scg}

\subsection{General framework}

The Lagrangian of SCG is built of metric variables and is only invariant under 3-dimensional spatial diffeomorphisms, which breaks time diffeomorphism. The action of such theories can be constructed in terms of scalars under spatial diffeomorphism. Consequently, it is natural and convenient to employ the Arnowitt-Deser-Misner (ADM) variables, which are based on the foliation structure of spacetime.

The general action for SCG, which extends the original proposal in \cite{Gao:2014soa,Gao:2019lpz} by allowing higher-order temporal derivatives, takes the form
\begin{equation}
	S=\int \mathrm{d}t\mathrm{d}^{3}x\,N\sqrt{h}\mathcal{L}\left(t,N,h_{ij},{}^{3}\!R_{ij},\nabla_{i},\pounds_{\bm{u}},\varepsilon_{ijk}\right), \label{SCG_ori}
\end{equation}
where $N$ is the lapse function, $h_{ij}$ is the 3-dimensional spatial metric, $R_{ij}$ is the 3-dimensional spatial Ricci tensor, $\nabla_{i}$ is the covariant derivative compatible with $h_{ij}$, $\bm{u}$ is the normal vector of the spatial hypersurfaces\footnote{Here we use $\bm{u}$ (instead of $\bm{n}$) to emphasize the scalar field $\phi$ specifying the hypersurfaces coincides with the time coordinate, which corresponds to the so-called unitary gauge.}. In order to introduce the parity-violating terms, the Levi-Civita tensor $\varepsilon_{ijk}\equiv\sqrt{h}\epsilon_{ijk}$ with $\epsilon_{123}=1$ is included. Please note the shift vector $N_{i}$ should not be included explicitly, since it is not a genuine geometrical quantity characterizing the foliation structure but merely encodes the gauge degrees of freedom of spatial diffeomorphism.

The formulation in terms of the ADM variables and the spacetime decomposition makes the kinetic terms be introduced in a natural and transparent manner. In particular, the separation of the temporal and spatial derivatives in the SCG allows us to focus on higher-order time derivatives, which could possibly introduce unwanted ghost-like or unwanted degrees of freedom without any restrictions. As mentioned above, the basic variables in SCG are the lapse function $N$ and the spatial metric $h_{ij}$. In the original proposal of \cite{Gao:2014soa,Gao:2014fra}, only the extrinsic curvature $K_{ij}$, which is the kinetic term for $h_{ij}$, is included in the Lagrangian. However, since time diffeomorphism is broken in SCG, $N$ should not be treated as an auxiliary field anymore. In fact, $N$ and $h_{ij}$ are independent and should be treated on equal footing, and thus the temporal derivatives of both variables should enter the Lagrangian through the Lie derivatives $\pounds_{\bm{u}}$, even up to higher orders.

Following \cite{Gao:2018znj,Gao:2019lpz}, the velocity of the lapse function $N$, namely $\dot{N}$, enters in the Lagrangian through 
	\begin{equation}
	F\equiv \frac{1}{N}\pounds_{\bm{u}}N=\frac{1}{N^{2}}\left(\dot{N}-N^{i}\nabla_{i}N\right)\,,  \label{Fdef}
	\end{equation}
with a dot denoting the time derivative $\partial_{t}$. The velocity of the spatial metric $h_{ij}$, namely $\dot{h}_{ij}$, enters the Lagrangian through the extrinsic curvature $K_{ij}$ defined by\footnote{Throughout this work, Lie derivatives acting on spatial tensors are understood as shorthands of the spatial components of the 4-dimension quantities. For example, $\pounds_{\bm{u}}h_{ij}\equiv e_{\phantom{a}i}^{a}e_{\phantom{b}j}^{b}\pounds_{\bm{u}}h_{ab}$, where $e^{a}_{\phantom{a}i}$ is the spatial components of the general basis $\bm{e}^{a}$.}
	\begin{equation}
	K_{ij}=\frac{1}{2}\pounds_{\bm{u}}h_{ij}=\frac{1}{2N}\left(\dot{h}_{ij}-\nabla_{j}N_{i}-\nabla_{i}N_{j}\right)\, .
	\end{equation}
The resulting action is given by
\begin{equation}
S=\int \mathrm{d}t\mathrm{d}^{3}x\,N\sqrt{h}\mathcal{L}\left(t,N,h_{ij},F,K_{ij},{}^{3}\!R_{ij},\nabla_{i},\varepsilon_{ijk}\right)\:. \label{SCG1}
\end{equation}
In the action \eqref{SCG1}, $N$ and $h_{ij}$ both act as dynamical variables, with no higher-order time derivatives appearing in the action. Other derivatives are purely spatial, which automatically evade the unwanted ghostlike mode. Generally speaking, there are four dynamical degrees of freedom (DoFs), consisting of two tensor and two scalar DoFs. Extensive research has been conducted on the degeneracy conditions under which the number of dynamical DoFs reduces to three \cite{Gao:2018znj,Gao:2019lpz} (and to two \cite{Lin:2020nro}, i.e., without any scalar DoF)\footnote{The time derivative of the lapse function is also discussed in \cite{Langlois:2017mxy}.}. In other words, additional conditions must be applied to ensure that only a single scalar DoF is present. It has been demonstrated that, up to the quadratic order in $K_{ij}$ and $F$, the resulting theory can be reduced to the one without $F$ via disformal transformations.

In principle, higher-order temporal derivatives, such as $\pounds_{\bm{u}}^2 N$ and $\pounds_{\bm{u}}^2 h_{ij}$ can also be considered, although generally they will introduce unwanted or ghost-like DoFs. Nevertheless, in this work, we will consider the Lie derivative of $N$ in terms of $F$ defined in (\ref{Fdef}) and the second-order Lie derivative of $h_{ij}$ in terms of $\pounds_{\bm{u}} K_{ij}$, which will be discussed in detail in Sec. \ref{sec:pvscg}.

\subsection{SCG with non-dynamical lapse function} \label{subsec:ndlapse}

The SCG theory proposed in \cite{Gao:2014soa,Gao:2014fra} (with parity-violating extension) is described by the action
\begin{equation}
	S=\int \mathrm{d}t\mathrm{d}^{3}x\,N\sqrt{h}\mathcal{L}\left(t,N,h_{ij},K_{ij},{}^{3}\!R_{ij},\nabla_{i},\varepsilon_{ijk}\right),
	\label{SCG2}
\end{equation}
in which the only time derivative entering in the Lagrangian is encoded in $K_{ij}$.
The theory \eqref{SCG2} has been proved to possess 3 DoFs through a Hamiltonian analysis \cite{Gao:2014fra}\footnote{The presence of $\varepsilon_{ijk}$ does not alter the constraint structure of the theory.}.

For the purpose of obtaining concrete 3-DoF ghost-free models, SCG Lagrangians of the polynomial type are considered and investigated in \cite{Gao:2020yzr}.
In order to get general theoretical forms while keeping the number of monomials finite, additional restrictions are necessary to exhaust all possible monomials.
In \cite{Gao:2020yzr}, it is assumed that the total number of derivatives does not exceed 4.
As we will see later, this implies that the Riemann curvature tensor is up to the quadratic order, the derivatives (both spatial and Lie derivatives) of $N$ and $h_{ij}$ are up to second order.
Degeneracy conditions reducing 3-DoF SCG theories to 2-DoF SCG theories can be found in \cite{Gao:2019twq,Hu:2021yaq} by different approaches.
It has been demonstrated that, under specific conditions, this type of SCG can exhibit only two tensorial DoFs without propagating any scalar mode.

It is necessary to make a classification of the possible SCG monomials due to their large number.
We will follow the same conventions and classification as in \cite{Gao:2020qxy,Gao:2020juc}.
By the Stueckelberg trick, an SCG term can be mapped to a corresponding generally covariant scalar-tensor (GST) term. Accordingly, we can assign each SCG term the set of integers ($c_{0}$; $d_{2}$, $d_{3}$) to characterize the GST combination, where $c_{0}$ is the number of Riemann curvature tensors, $d_{2}$, $d_{3}$ are the numbers of the second and the third covariant derivatives of $\phi$, respectively\footnote{Up to $d=4$ with $d$ being the total number of derivatives, higher-order integers such as $c_1$ and $d_{4}$ etc., are not needed, since the corresponding SCG monomials can always be reduced.}.
Schematically, we can write
\begin{align}
	K_{ij} & \sim a_{i}\sim(0;1,0),\\
	{}^{3}\!R_{ij} & \sim(1;0,0),\\
	\nabla_{k}K_{ij} & \sim\nabla_{j}a_{i}\sim(0;0,1).
\end{align}
Thus, the total number of derivatives in each monomial is given by
\begin{align}
	d=\sum_{n=0}\left[(n+2)c_{n}+(n+1)d_{n+2}\right].
\end{align}
Since the resulting GST terms would arise only from the combinations of GST monomials of the same values of $d$, the number $d$ can be used as a generic label characterizing monomials of both GST and SCG, which makes the correspondences more transparent. In this work, we consider monomials with $d\leq4$.
Even considering higher-order operators, one would find that no new independent monomials are present in each category, as long as the value of the total number of derivatives $d$ is not larger than 4.

Based on (\ref{SCG2}), a specific ghost-free SCG Lagrangian of the polynomial type with parity violation was proposed in \cite{Gao:2020yzr}. The parity-violating terms are denoted by $\tilde{\mathcal{L}}^{(3)}$ for $d=3$ and $\tilde{\mathcal{L}}^{(4)}$ for $d=4$, which are given by
	\begin{equation}
		\mathcal{\tilde{L}}^{(3)}= c_{1}^{(0;1,1)}\varepsilon_{ijk}K_{l}^{i}\nabla^{j}K^{kl},
	\end{equation}
and 
\begin{eqnarray}
\mathcal{\tilde{L}}^{(4)} & = & c_{1}^{(0;2,1)}\varepsilon_{ijk}K^{im}K^{jn}\nabla_{m}K_{n}^{k}+c_{2}^{(0;2,1)}\varepsilon_{ijk}K^{mn}K_{m}^{i}\nabla^{j}K_{n}^{k} \nonumber \\ 
 & & +c_{3}^{(0;2,1)}\varepsilon_{ijk}K_{l}^{i}a^{j}\nabla^{k}a^{l}+c_{4}^{(0;2,1)}\varepsilon_{ijk}K_{l}^{i}\nabla^{j}K^{kl}K \nonumber \\
&  & +c_{1}^{(1;2,0)}\varepsilon_{ijk}{}^{3}\!R_{l}^{i}K^{jl}a^{k}+c_{1}^{(1;0,1)}\varepsilon_{ijk}{}^{3}\!R_{l}^{i}\nabla^{j}K^{kl},
\end{eqnarray}
respectively.
The coefficients $c_{m}^{(c_{0};d_{2},d_{3})}$ are generally functions of $t$ and $N$. 

In the next section, we will extend the above Lagrangian by introducing new parity-violating terms involving Lie derivatives of the lapse function and the extrinsic curvature.

\section{Parity-violating monomials in SCG}\label{sec:pvscg}

In this section, by considering novel SCG operators involving higher-order derivatives, we will construct and classify the parity-violating SCG monomials in a systematic way under the restriction $d\leq 4$, with $d$ being the total number of derivatives.

\subsection{Building blocks}

First of all, we need to extend the set of operators for our purpose.
Fundamental geometrical quantities, including the lapse function $N$, the spatial metric $h_{ij}$, the acceleration $a_{i}$, the extrinsic curvature $K_{ij}$, and the spatial Ricci tensor ${}^3\!R_{ij}$, along with their spatial derivatives, are involved as the basic ingredients.
It should be noted that these quantities are associated with the foliation defined by the scalar field once the time coordinate is fixed.
Moreover, we will introduce higher-order derivatives of the lapse function $N$ and the spatial metric $h_{ij}$.
Precisely, spatial and Lie derivatives of $K_{ij}$, $a_{i}$, and $F$, i.e., $\nabla_{k}K_{ij}$, $\nabla_{i} a_{j}$, $\nabla_i F$, $\pounds_{\bm{u}}K_{ij}$, and $\pounds_{\bm{u}}F$, as well as $\pounds_{\bm{u}}a_{i}$, will be taken into account in our construction.
The reason for introducing these higher-order derivative operators is not only because they would extend the SCG construction, but also because they naturally (or necessarily) arise in the unitary gauge for GST monomials up to $d=4$.

When evaluating Lie derivatives with respect to the normal vector $\bm{u}$, it is necessary to transform the expressions into a generally covariant form.
Once it is established that these operators are indeed spatial tensors, we can then take their spatial components with indices $i,j,\cdots$.
For example, the Lie derivative of the extrinsic curvature with lower indices is a spatial tensor since $n^{b}\pounds_{\bm{u}}K_{ab}=n^{b}\pounds_{\bm{u}}K_{ba}=0$.
More explicitly, we have
\begin{equation}
	\pounds_{\bm{u}}K_{ij}=\frac{1}{N}\left[\dot{K}_{ij}-\left(N^{k}\nabla_{k}K_{ij}+K_{kj}\nabla_{i}N^{k}+K_{ki}\nabla_{j}N^{k}\right)\right],
\end{equation}
which involves the second order time derivative of $h_{ij}$ through $\dot{K}_{ij}$.
On the other hand, one can show that Lie derivative of the acceleration can be reduced, 
\begin{equation}
	\pounds_{\bm{u}}a_{i}=\pounds_{\bm{u}}\left(\frac{\nabla_{i}N}{N}\right)=\nabla_{i}F+Fa_{i},
\end{equation}
which implies that $\pounds_{\bm{u}}a_{i}$ is not an independent operator. 
A subtle point arises when taking Lie derivatives of spatial tensors with upper indices\footnote{For instance, Lie derivatives of the extrinsic curvature with mixed or upper indices are given by
	\begin{eqnarray*}
		\pounds_{\bm{u}}K_{c}^{d} & = & h^{da}\pounds_{\bm{u}}K_{ac}-2K_{c}^{a}K_{a}^{d}+a^{a}K_{ac}n^{d}\neq h^{da}\pounds_{\bm{u}}K_{ac},\\
		\pounds_{\bm{u}}K^{ab} & = & h^{ad}h^{bc}\pounds_{\bm{u}}K_{dc}-4K^{ae}K_{e}^{b}+a_{e}\left(K^{be}n^{a}+K^{ae}n^{b}\right)\neq h^{bc}h^{ad}\pounds_{\bm{u}}K_{dc},
	\end{eqnarray*}
which are not spatial tensors neither.}. 
To avoid confusion, we only use the Lie derivative of spatial tensors with lower indices in SCG. Lie derivatives of spatial tensors with mixed or upper indices are understood as merely shorthands.
For example, only the Lie derivative $\pounds_{\bm{u}}K_{ij}$ (spatial component of $\pounds_{\bm{u}}K_{ab}$) is defined in SCG, while $\pounds_{\bm{u}}K_{j}^{i}$ is just a shorthand that stands for
\begin{equation}
	\pounds_{\bm{u}}K_{j}^{i}\equiv h^{ik}\pounds_{\bm{u}}K_{kj}=h^{ik}e_{\phantom{a}k}^{c}e_{\phantom{b}j}^{b}\pounds_{\bm{u}}K_{cb}=e_{a}^{\phantom{a}i}e_{\phantom{b}j}^{b}h^{ac}\pounds_{\bm{u}}K_{cb}.
\end{equation}
Similarly, $\pounds_{\bm{u}}K^{ij}\equiv h^{ik}h^{jl}\pounds_{\bm{u}}K_{kl}$, etc.
Note we also have a useful relation for the spatial derivative of the acceleration $\nabla_i a_j$:
\begin{equation}
	\nabla_{i}a_{j}=\nabla_{i}\nabla_{j}\ln N=\nabla_{j}\nabla_{i}\ln N=\nabla_{j}a_{i},
\end{equation}
which implies that its indices are symmetric.

Moreover, up to $d=4$, SCG monomials involving third-order or fourth-order derivatives can be reduced to equivalent combinations of the same order of $d$ but with lower-order operators through integration by parts. This allows for the construction of a simplified and complete set of basis monomials, facilitating analysis and computations.

Finally, we have a complete set of operators for our purpose. The lapse function $N$ and the induced metric $h_{ij}$ can be regarded as
the 0-th order operator. The independent operators up to the second order in derivatives are shown in Table \ref{SCG operators}.
\begin{table}[h]
\begin{centering}
\renewcommand\arraystretch{1.6}
\begin{tabular}{|c|ccc|ccc|}
\hline 
0 order &  $\quad h_{ij}$ &  &  & $\quad N$ &  & \tabularnewline
\hline 
\hline 
1st order & $\quad K_{ij}$ &  &  & $\quad F,$ &  & $a_{k}\quad \quad $ \tabularnewline
\hline 
2nd order & $\quad \quad \pounds_{\bm{u}}K_{ij}$, & $\nabla_{k}K_{ij},$ & $^{3}\!R_{ij}\quad \quad $ & $\quad \quad \pounds_{\bm{u}}F,$ & $\nabla_{k}F,$ & $\quad \nabla_{i}a_{j}\quad \quad $\tabularnewline
\hline 
$\dots$  & $\dots$ &  &  & $\dots$ &  &  \tabularnewline
\hline 
\end{tabular}
\par\end{centering}
\caption{Building blocks in SCG.}
\label{SCG operators}
\end{table}
As we have mentioned above, up to $d=4$ we do not need to consider operators involving derivatives higher than the second order (acting on the ADM variables), such as $\pounds_{\bm{u}}{}^{3}\!R_{ij}$, $\pounds_{\bm{u}}^{2}K_{ij}$, $\nabla^{i}\nabla^{j}K_{ij}$, etc. 

\subsection{Complete basis for the parity-violating monomials}

To identify all the parity-violating terms constructed by these building blocks, we divide the entire process into two steps.
The first step involves determining all the possible types of combinations of the given order. In this step, we do not distinguish various contractions from each other. For notational simplicity, we use the initial letters to represent each geometric quantity, such as $a$ for $a_{k}$, $K$ for $K_{ij}$, $R$ for ${}^{3}\!R_{ij}$, $\pounds K$ for $\pounds_{\bm{u}}K_{ij}$, $\nabla K$ for $\nabla_{k}K_{ij}$, etc.
In the second step, we systematically explore all the possible contractions and identify all the independent monomials.
Note that the spatial Levi-Civita tensor $\varepsilon_{ijk}$ is necessary in order to construct monomials with parity violation.

\subsubsection{$d=3$}

There are no parity-violating monomials in the cases of $d = 0, 1, 2$. So we start from the case of $d=3$.
In the case of $d=3$, we have two types of operator combinations, schematically denoted by $[1+2]$ and $[1+1+1]$.
Here the integers denote orders of derivatives in each building block (operator). For example, a monomial built from contracting $F$ and $\nabla_iF$ will be of $[1+2]$ type, since $F$ contains the first-order derivative and $\nabla_iF$ contains the second-order derivative of $N$, respectively.

For the type of $[1+2]$, besides the combinations $[K\nabla K]$, $[a\nabla a]$, and $[aR]$ that have been considered in \cite{Gao:2020yzr,Gao:2020qxy}, we have the following new combinations
\begin{equation*}
	[F\nabla F],\quad [a\pounds F],\quad [K\nabla F], \quad [F\nabla K],\quad [a\pounds K],
\end{equation*}
due to the presence of $F$ and $\pounds_{\bm{u}}K_{ij}$.
Due to the (anti)symmetry of indices, no viable monomials arise from these parity-violating combinations except $[K\nabla K]$. Therefore, we refer to these cases as being ``empty'' and can disregard them in the subsequent step. The other type of operator combination, i.e., $[1+1+1]$, contains the combination $[aaa]$ as well as the new ones\footnote{Since we are considering parity-violating monomials, the epsilon tensor $\varepsilon_{ijk}$ must be present. Therefore, we only need to consider combinations of operators with an odd number of indices.}
\begin{equation*}
	[aKF],\quad [aFF].
\end{equation*}
Clearly, no viable parity-violating monomials emerge from these combinations.

As a result, it becomes evident that the only non-empty parity-violating combination is $[K\nabla K]$. Moreover, the only independent contraction within this combination is:
\begin{equation}
	[K\nabla K]:\quad \varepsilon_{ijk}K_{l}^{i}\nabla^{j}K^{kl},
\end{equation}
which has already been studied in \cite{Gao:2020yzr,Gao:2020qxy}.
In brief, in the case of $d=3$, despite the introduction of new operators $F$ and $\pounds_{\bm{u}}K_{ij}$, no new parity-violating SCG monomial emerges.

\subsubsection{$d=4$}

In the case of $d=4$, we have three different types of operator combinations: $[2+2]$, $[1+1+2]$, and $[1+1+1+1]$. Based on the same notation, we express all the possible combinations of each type.

For the type of $[1+1+1+1]$, besides the combinations $[aKKK]$, $[aaaK]$, we have the following new combinations due to the presence of $F$,
\begin{equation*}
[KKaF],\quad [aaaF],\quad [KaFF],\quad [aFFF]\:.
\end{equation*}
No viable parity-violating monomials exist in these types of combinations.

For the type of $[1+1+2]$, there are viable parity-violating monomials from the combination types
	\begin{equation}
		 [aK\nabla a],\quad [aa\nabla K],\quad [KK\nabla K], \quad [aKR],
	\end{equation}
which have been considered in \cite{Gao:2020yzr,Gao:2020qxy}.
In addition, there are new viable parity-violating monomials corresponding to the combinations
\begin{equation}
	[aK \pounds K],\quad [FK\nabla K].
\end{equation}
While the following combinations
\begin{align*}
&[Ka\pounds F],\quad [Fa\pounds K],\quad [Fa\pounds F],\quad [Fa\nabla a],\quad [FaR],\\
  &[FF\nabla K],\quad [KK\nabla F],\quad [KF\nabla F],\quad [FF\nabla F],\quad [aa\nabla F]\:,
\end{align*}
yield no viable monomials.

For the type of $[2+2]$, besides the combinations $[R\nabla K]$ and $[\nabla a \nabla K]$, we have one new combination
\begin{equation}
	[\pounds K\nabla K],
\end{equation}
with empty cases 
\[
[\pounds F\nabla F],\quad [\pounds K\nabla F],\quad [\pounds F\nabla K],\quad [R\nabla F],\quad [\nabla a\nabla F]\:.
\] 

In summary, due to the presence of $F$ and $\pounds_{\bm{u}}K_{ij}$, there arise 3 new parity-violating SCG monomials in the case of $d=4$, which are
\begin{eqnarray}
[FK\nabla K] & : &\quad \{\varepsilon_{ijk}FK_{l}^{i}\nabla^{j}K^{kl}\},\\{}
[aK\pounds K] & : &\quad \{\varepsilon_{ijk}a^{i}K^{jl}\pounds_{\bm{u}}K_{l}^{k}\},\\{}
[\pounds K\nabla K] & : &\quad \{\varepsilon_{ijk}\pounds_{\bm{u}}K^{li}\nabla^{j}K_{l}^{k}\}.
\end{eqnarray}

At this point, we note that not all these terms are independent. One can check that the last term, i.e., $\varepsilon_{ijk}\pounds_{\bm{u}}K^{li}\nabla^{j}K_{l}^{k}$, is not independent and can be expressed by linear combinations of the other terms up to total derivatives.
The details can be found in Appendix \ref{app:rel}.
Therefore, in the following, we do not need to consider $\varepsilon_{ijk}\pounds_{\bm{u}}K^{li}\nabla^{j}K_{l}^{k}$.

\begin{table}[h]
	\begin{centering}
		\renewcommand\arraystretch{1.6}
		\resizebox{\linewidth}{!}{
			\begin{tabular}{|c|c|c|c|c|c|}
				\hline 
				$d$ & Category & Form & Irreducible & Reducible & Number\tabularnewline
				\hline 
				\hline 
				3 & $\left(0; 1,1\right)$ & $[K\nabla K]$ & $\varepsilon_{ijk}K_{l}^{i}\nabla^{j}K^{kl}$ & $-$ & 1 \tabularnewline
				\hline 
				&$\left(0; 0,2\right)$ & $[\pounds K \nabla K]$ & $-$ & $\varepsilon_{ijk}\pounds_{\bm{u}}K^{li}\nabla^{j}K_{l}^{k}$ & 0 \tabularnewline
				\cline{3-5} \cline{4-5} \cline{5-5} 
				4 &  & $[\nabla a \nabla K]$ & $-$ & $\varepsilon_{ijk}\nabla^{l}a^{i}\nabla^{j}K_{l}^{k}$ & \tabularnewline
				\cline{2-6} \cline{3-6} \cline{4-6} \cline{5-6} \cline{6-6} 
				& $\left(0; 2,1\right)$ & $[aK \nabla a]$ & $\varepsilon_{ijk}a^{i}\nabla^{l}a^{j}K_{l}^{k}$ & $-$ & \tabularnewline
				\cline{4-5} \cline{5-5} 
				&  & $[aa\nabla K]$ & {$-$} & {$\varepsilon_{ijk}a^{l}a^{i}\nabla^{j}K_{l}^{k}$} & \tabularnewline
				\cline{4-5} \cline{5-5} 
				&  & $[KK\nabla K]$ & $\varepsilon_{ijk}K_{l}^{i}K_{m}^{j}\nabla^{m}K^{kl},\:\varepsilon_{ijk}K_{l}^{i}K^{ml}\nabla^{j}K_{m}^{k},\:\varepsilon_{ijk}KK_{l}^{i}\nabla^{j}K^{kl}$ & $-$ & 6\tabularnewline
				\cline{4-5} \cline{5-5} 
				&  & $[aK\pounds K]$ & $\varepsilon_{ijk}a^{i}K^{jl}\pounds_{\bm{u}}K_{l}^{k}$ & $-$ & \tabularnewline
				\cline{4-5} \cline{5-5} 
				&  & $[FK\nabla K]$ & $\varepsilon_{ijk}FK_{l}^{i}\nabla^{j}K^{kl}$ & $-$ & \tabularnewline
				\cline{2-6} \cline{3-6} \cline{4-6} \cline{5-6} \cline{6-6} 
				& $\left(1; 2,0\right)$ & $[RKa]$ & $\text{\ensuremath{\varepsilon_{ijk}}}a^{i}K^{jl}{}^{3}\!R_{l}^{k}$ & $-$ & 1\tabularnewline
				\cline{2-6} \cline{3-6} \cline{4-6} \cline{5-6} \cline{6-6} 
				&$\left(1; 0,1\right)$ & $[R\nabla K]$ & $\text{\ensuremath{\varepsilon_{ijk}}}{}^{3}\!R^{li}\nabla^{j}K_{l}^{k}$ & $-$ & 1\tabularnewline
				\hline 
			\end{tabular}
		}
		\par\end{centering}
	\caption{Classification of the parity-violating SCG monomials.}
	\label{SCG_pv_mono}
\end{table}

We list all the irreducible and reducible parity-violating SCG monomials in Table \ref{SCG_pv_mono}.
In the case of $d=3$, there is only one parity-violating monomial. In the case of $d=4$, there are 8 independent parity-violating monomials.
Following the terminology developed in \cite{Gao:2020qxy}, the complete basis of parity-violating SCG polynomials for $d=3$ is thus
\begin{equation}
	\mathcal{Z}_{3} = \left\{ \varepsilon_{ijk}K_{l}^{i}\nabla^{j}K^{kl}\right\} , \label{Z3}
\end{equation}
which is composed of a single unfactorizable and irreducible monomial. In other words, we have $\dim\left(\mathcal{Z}_{3}\right)=1$.
The ``enlarged'' basis for the parity-violating SCG polynomials for $d=4$ is given by 
\begin{eqnarray}
	\mathcal{Z}_{4} & = & \Big\{\varepsilon_{ijk}{}^{3}\!R_{l}^{i}K^{jl}a^{k},\varepsilon_{ijk}K^{im}K^{jn}\nabla_{m}K_{n}^{k},\varepsilon_{ijk}K^{mn}K_{m}^{i}\nabla^{j}K_{n}^{k},\varepsilon_{ijk}K_{l}^{i}a^{j}\nabla^{k}a^{l},\nonumber \\
	&  & \varepsilon_{ijk}{}^{3}\!R_{l}^{i}\nabla^{j}K^{kl},\varepsilon_{ijk}KK_{l}^{i}\nabla^{j}K^{kl},\varepsilon_{ijk}a^{i}K^{jl}\pounds_{\bm{u}}K_{l}^{k},\varepsilon_{ijk}FK_{l}^{i}\nabla^{j}K^{kl}\Big\}, \label{Z4}
\end{eqnarray}
with $\dim\left(\mathcal{Z}_{4}\right)=8$.
Note that the monomial in (\ref{Z3}), as well as the first six monomials in (\ref{Z4}), contain no $F$ or $\pounds_{\bm{u}}K_{ij}$ and thus are ghost-free. However, the last two monomials in (\ref{Z4}) involve $F$ or $\pounds_{\bm{u}}K_{ij}$ (although in a linear manner) and thus possibly suffer from the Ostrogradsky ghost even in the SCG.

\section{Ghost-free scalar-tensor theories with parity violation} \label{sec:gfgst}

The GST monomials are systematically classified, and the complete basis for GST polynomials up to $d=4$ is derived in \cite{Gao:2020juc,Gao:2020yzr}. We refer to Appendix \ref{app:GSTmono} for the explicit expressions of the parity-violating GST monomials, which we denote $F_{n}^{(c_0;d_2,d_3)}$ for short.
Our purpose is to find the ghost-free combinations of these GST monomials. To this end, we will first derive the corresponding expressions of these parity-violating monomials in the unitary gauge, which explicitly show the dangerous terms $\sim \pounds_{\bm{u}}K_{ij}$ and $F$, thus giving us guidance on how to build the ghost-free combinations.

Before proceeding, we would like to clarify the notation in this work. For our purpose, we frequently switch among different but equivalent formulations of the theory.
First, we refer to SCG, in which all the quantities are spatial tensors with spatial indices $i,j,k,\cdots$. For example, the actions (\ref{SCG_ori}) and (\ref{SCG1}) are of the SCG form.
Sometimes it is convenient or even necessary to work with the generally covariant correspondence of the SCG terms.
For example, for the extrinsic curvature, we may work with $K_{ab}$ instead of $K_{ij}$, where the former is understood as $K_{ab} = \frac{1}{2} \pounds_{\bm{u}} h_{ab}$ with $u_{a} = -N \nabla_{a} \phi$ and $h_{ab} = g_{ab} + u_{a}u_{b}$.
In principle, it is essentially a generally covariant scalar-tensor theory term, but ``wrapped'' in terms of hypersurface geometrical quantities.
Following the strategy in \cite{Hu:2021bbo}, we refer to expressions in such a form as the ``unitary gauge'' or ``u.g.'' form for short\footnote{We emphasize that in the literature ``unitary gauge'' is often referred to as ``SCG'' in our terminology, which is written in a concrete coordinate system. For our purpose, we refer to the unitary gauge as merely a special choice of the normal vector $n_{a} \rightarrow u_{a}\equiv -N \nabla_{a} \phi$, which has nothing to do with any specific coordinates. See the discussion in \cite{Hu:2021bbo} for details.}.

\subsection{Decomposition in the unitary gauge} \label{subsec:dec}

First of all, there are 4 parity-violating GST monomials that do not contain higher-order time derivatives in the unitary gauge, i.e., without the dangerous terms $\pounds_{\bm{u}}K_{ab}$ or $F$, which are
	\begin{equation}
		F_{1}^{(1;1,0)}\xlongequal{\text{u.g.}}2\varepsilon_{abc}K^{ad}\mathrm{D}^{b}K_{d}^{c}, \label{F11011dec}
	\end{equation}
	\begin{equation}
		F_{6}^{(0;2,1)}\xlongequal{\text{u.g.}}-\varepsilon_{abc}K_{d}^{a}a^{b}\mathrm{D}^{c}a^{d},
	\end{equation}
	\begin{equation}
		F_{2}^{(1;2,0)}\xlongequal{\mathrm{u.g.}}\varepsilon_{abc}\left(-2a^{a}a^{d}\mathrm{D}^{c}K_{d}^{b}+2K_{d}^{a}K^{de}\mathrm{D}^{c}K_{e}^{b}\right),\label{F1202dec}
	\end{equation}
and
	\begin{equation}
		F_{7}^{(1;2,0)}\xlongequal{\mathrm{u.g.}}2\varepsilon_{abc}a^{a}a^{d}\mathrm{D}^{c}K_{d}^{b},\label{F1207dec}
	\end{equation}
with $\sigma\equiv \sqrt{-\phi_{a}\phi^{a}}$. Here and in the following, we define
	\begin{equation}
		\varepsilon_{abc} \equiv u^{d} \varepsilon_{dabc}, \label{eps3def}
	\end{equation}
and ``u.g.'' denotes equality in the unitary gauge.
In deriving the above expressions, no integration by parts has been performed.
As a result, we can conclude that these 4 GST monomials are automatically free of the Ostrogradsky ghost in the unitary gauge.

Before proceeding, note that the SCG monomials in the decomposition of $F^{(1;2,0)}_{2}$ and $F^{(1,2,0)}_{7}$ in (\ref{F1202dec}) and (\ref{F1207dec}) are not precisely the SCG monomials chosen in the complete basis (\ref{Z4}).
Nevertheless, they can be recast as linear combinations of the SCG monomials in the complete basis up to total derivatives.
The relevant expressions are given in (\ref{F1202ibp}) and (\ref{F1207ibp}).
We also refer to Appendix \ref{app:ibp} for more detailed integrations by parts we have used in deriving the above and in the following expressions.
At this point, note that the expressions of $F_{6}^{(0;2,1)}$ and $F_{7}^{(1;2,0)}$ in the unitary gauge differ by a total derivative term. Indeed, this also happens for their general covariant expressions (\ref{app:F0216red}), which implies that although $F_{6}^{(0;2,1)}$ is algebraically independent, it is not independent in the sense of integration by parts.

According to Appendix \ref{app:GSTmono}, there are in total 10 parity-violating GST monomials in the $(1;2,0)$ category. Besides $F_{2}^{(1;2,0)}$ and $F_{7}^{(1;2,0)}$ mentioned above, the remaining 8 monomials in the $(1;2,0)$ category will contribute to terms linear in $F$ and/or $\pounds_{\bm{u}}K_{ij}$ in the unitary gauge.
The following three monomials would contribute to the $[FK\nabla K]$ term:
	\begin{equation}
		F_{4}^{(1;2,0)}\xlongequal{\mathrm{u.g.}}\varepsilon_{abc}\left(2a^{a}K^{db}\,^{3}\!R_{d}^{c}-2a^{a}a^{d}\mathrm{D}^{c}K_{d}^{b}-2F\mathrm{D}^{a}K_{d}^{b}K^{cd}\right), \label{F1204dec}
	\end{equation}
	\begin{equation}
		F_{9}^{(1;2,0)} \xlongequal{\text{u.g.}}  \varepsilon_{abc}\left(2FK^{ad}\mathrm{D}^{b}K_{d}^{c}-2KK^{ea}\mathrm{D}^{b}K_{e}^{c}\right),
	\end{equation}
	and
	\begin{equation}
		F_{10}^{(1;2,0)} \xlongequal{\text{u.g.}}  -2\varepsilon_{abc}FK^{ad}\mathrm{D}^{b}K_{d}^{c}.
	\end{equation}
SCG monomials in the decomposition of $F^{(1;2,0)}_{4}$ in (\ref{F1204dec}) have not been in the form of the complete basis (\ref{Z4}). Similar to $F^{(1;2,0)}_{2}$ and $F^{(1,2,0)}_{7}$ above, it can be recast as a linear combination of monomials in the complete basis up to a total derivative, which is given in (\ref{F1204ibp}).
The following three monomials contribute to the $[aK\pounds K]$ term:
	\begin{equation}
		F_{5}^{(1;2,0)} \xlongequal{\text{u.g.}}  \varepsilon_{abc}\left(-K^{ad}K^{be}\mathrm{D}_{e}K_{d}^{c}-K_{d}^{a}a^{b}\mathrm{D}^{c}a^{d}-a^{a}K^{db}h^{ec}\pounds_{\boldsymbol{u}}K_{de}\right),
	\end{equation}
	\begin{equation}
		F_{6}^{(1;2,0)} \xlongequal{\text{u.g.}}  \varepsilon_{abc}\left(-\,{}^{3}\!R_{d}^{a}K^{db}a^{c}+K_{d}^{a}a^{b}\mathrm{D}^{c}a^{d}+a^{a}K^{db}h^{ec}\pounds_{\boldsymbol{u}}K_{de}\right),
	\end{equation}
and
	\begin{equation}
		F_{8}^{(1;2,0)} \xlongequal{\text{u.g.}}  \varepsilon_{abc}\left(-K_{d}^{a}a^{b}\mathrm{D}^{c}a^{d}-a^{a}K^{db}h^{ec}\pounds_{\boldsymbol{u}}K_{de}\right).
	\end{equation}
Finally, the following two monomials would contribute to both $[FK\nabla K]$ and $[aK\pounds K]$ terms: 
	\begin{eqnarray}
		F_{1}^{(1;2,0)} & \xlongequal{\mathrm{u.g.}} & 4\varepsilon_{abc}\Big(a^{a}K^{db}\,^{3}\!R_{d}^{c}-a^{a}K_{d}^{c}\mathrm{D}^{b}a^{d}-a^{a}a^{d}\mathrm{D}^{c}K_{d}^{b}-F\mathrm{D}^{a}K_{d}^{b}K^{cd}\nonumber \\
		&  & \qquad-K^{ad}K^{be}\mathrm{D}_{e}K_{d}^{c}-a^{a}K^{db}h^{ec}\pounds_{\boldsymbol{u}}K_{de}\Big), \label{F1201dec}
	\end{eqnarray}
	\begin{eqnarray}
		F_{3}^{(1;2,0)} & \xlongequal{\mathrm{u.g.}} & 2\varepsilon_{abc}\left(a^{a}K_{d}^{c}\mathrm{D}^{b}a^{d}+a^{a}a^{d}\mathrm{D}^{c}K_{d}^{b}+F\mathrm{D}^{a}K_{d}^{b}K^{cd}+a^{a}K^{db}h^{ec}\pounds_{\boldsymbol{u}}K_{de}\right).\qquad \label{F1203dec}
	\end{eqnarray}
Again, (\ref{F1201dec}) and (\ref{F1203dec}) can be recast as linear combinations of monomials in the complete basis through integrations by parts, which are given in (\ref{F1201ibp}) and (\ref{F1203ibp}), respectively.
Thus, within the $(1;2,0)$ category, in order to make the theory healthy in the unitary gauge, we obtain three special combinations that cancel both dangerous terms $[aK\pounds K]$ and $[FK\nabla K]$, such as $F_{1}^{(1;2,0)}-2F_{4}^{(1;2,0)}+4F_{6}^{(1;2,0)}$,$\quad F_{3}^{(1;2,0)}+F_{4}^{(1;2,0)}-2F_{6}^{(1;2,0)}$ and $F_{5}^{(1;2,0)}+F_{6}^{(1;2,0)}$. These combinations are totally healthy in the unitary gauge with any coefficient functions of $\phi$ and $X$. 

There are in total 5 monomials of the $(2;0,0)$ category. It is easy to show that all the monomials in this category would produce dangerous terms of the $[\pounds K\nabla K]$ form. We have
	\begin{eqnarray}
		F_{1}^{(2;0,0)} & \xlongequal{\mathrm{u.g.}} & 16\varepsilon_{abc}\Big(\mathrm{D}^{a}a^{d}\mathrm{D}^{c}K_{d}^{b}+a^{a}a^{d}\mathrm{D}^{c}K_{d}^{b}+\,{}^{3}\!R^{da}\mathrm{D}^{c}K_{d}^{b}+K_{d}^{a}K^{de}\mathrm{D}^{c}K_{e}^{b}\nonumber \\
		&  & \qquad-K^{da}K^{eb}\mathrm{D}_{e}K_{d}^{c}-h^{da}\mathrm{D}^{c}K^{eb}\pounds_{\boldsymbol{u}}K_{de}\Big), \label{F2001dec}
	\end{eqnarray}
	\begin{equation}
		F_{2}^{(2;0,0)}\xlongequal{\mathrm{u.g.}}8\varepsilon_{abc}\Big(\mathrm{D}^{a}a^{d}\mathrm{D}^{c}K_{d}^{b}+a^{a}a^{d}\mathrm{D}^{c}K_{d}^{b}+K_{d}^{a}K^{de}\mathrm{D}^{c}K_{e}^{b}-h^{da}\mathrm{D}^{c}K^{eb}\pounds_{\boldsymbol{u}}K_{de}\Big),
	\end{equation}
	\begin{eqnarray}
		F_{3}^{(2;0,0)} & \xlongequal{\mathrm{u.g.}} & 4\varepsilon_{abc}\Big(-\mathrm{D}^{a}a^{d}D^{c}K_{d}^{b}-{}^{3}\!R^{da}\mathrm{D}^{c}K_{d}^{b}-a^{a}a^{d}\mathrm{D}^{c}K_{d}^{b}\nonumber \\
		&  & \qquad-K_{d}{}^{a}K^{de}\mathrm{D}^{c}K_{e}^{b}+K^{da}K^{eb}\mathrm{D}_{e}K_{d}{}^{c}+h^{da}\mathrm{D}^{c}K^{eb}\pounds_{\boldsymbol{u}}K_{de}\Big),
	\end{eqnarray}
	\begin{eqnarray}
		F_{4}^{(2;0,0)} & \xlongequal{\mathrm{u.g.}} & 2\varepsilon_{abc}\Big(-\mathrm{D}^{a}a^{d}\mathrm{D}^{c}K_{d}^{b}+{}^{3}\!R^{da}\mathrm{D}^{c}K_{d}^{b}-a^{a}a^{d}\mathrm{D}^{c}K_{d}^{b}\nonumber \\
		&  & \qquad-2K_{d}^{a}K^{de}\mathrm{D}^{c}K_{e}^{b}+KK^{ea}\mathrm{D}^{c}K_{e}^{b}+h^{da}\mathrm{D}^{c}K^{eb}\pounds_{\boldsymbol{u}}K_{de}\Big),
	\end{eqnarray}
	\begin{equation}
		F_{5}^{(2;0,0)}\xlongequal{\mathrm{u.g.}}2\varepsilon_{abc}\Big(\mathrm{D}^{a}a^{d}\mathrm{D}^{c}K_{d}^{b}+a^{a}a^{d}\mathrm{D}^{c}K_{d}^{b}+K_{d}^{a}K^{de}\mathrm{D}^{c}K_{e}^{b}-h^{da}\mathrm{D}^{c}K^{eb}\pounds_{\boldsymbol{u}}K_{de}\Big). \label{F2005dec}
	\end{equation}
Again, (\ref{F2001dec})-(\ref{F2005dec}) can be recast as linear combinations of monomials in the complete basis through integrations by parts, which are given in (\ref{F2001ibp})-(\ref{F2005ibp}), respectively.
Thus, within the $(2;0,0)$ category, it is easy to find some special combinations without the $[\pounds K\nabla K]$ term, such as
$F_{2}^{(2;0,0)}-\frac{1}{2}F_{1}^{(2;0,0)}$,\,$F_{3}^{(2;0,0)}+\frac{1}{4}F_{1}^{(2;0,0)}$,\,$F_{4}^{(2;0,0)}+\frac{1}{8}F_{1}^{(2;0,0)}$ and $F_{5}^{(2;0,0)}-\frac{1}{8}F_{1}^{(2;0,0)}$.

\subsection{Ghost-free scalar-tensor polynomials} \label{subsec:gfst}

Based on the covariant scalar-tensor terms corresponding to the SCG terms developed in \cite{Gao:2020yzr}, the covariant correspondence of the SCG monomials, including the parity-violating monomials, has been shown in \cite{Gao:2020qxy}. However, in \cite{Gao:2020qxy}, only the matrices of the linear mapping are shown, instead of the explicit expressions for the covariant correspondence. Moreover, the higher-order Lie derivatives have not been taken into account in \cite{Gao:2020yzr}. In the following, we will show the explicit expressions for the covariant correspondence of the parity-violating SCG monomials up to $d=4$.

For $d=3$, the only parity-violating SCG monomial is given in (\ref{Z3}), of which the covariant correspondence is \cite{Crisostomi:2017ugk,Gao:2020yzr,Gao:2020qxy}
\begin{eqnarray}
	\varepsilon_{ijk}K_{l}^{i}\nabla^{j}K^{kl} & \rightarrow & \varepsilon^{abcd}u_{a}K_{b}^{e}\mathrm{D}_{c}K_{de}\nonumber \\
	& = & \frac{1}{2\left(2X\right)^{3/2}}\varepsilon^{abcd}R_{cdef}\phi_{a}\phi^{e}\phi_{b}^{f}\equiv\frac{1}{2}F_{1}^{(1;1,0)}. \label{d3cor}
\end{eqnarray}
(\ref{d3cor}) can be compared with (\ref{F11011dec}).

For $d=4$, the first 6 monomials in the complete basis (\ref{Z4}) do not involve $F$ or $\pounds_{u}K_{ij}$, the covariant correspondence of which is
\begin{eqnarray}
	\varepsilon_{ijk}{}^{3}\!R_{l}^{i}K^{jl}a^{k} & \rightarrow & \varepsilon^{abcd}u_{a}a_{b}K_{c}^{e}\,{}^{3}\!R_{ed}\nonumber \\
	& = & -\frac{1}{4X^{2}}\varepsilon^{abcd}R_{af}\phi_{b}\phi^{e}\phi_{ed}\phi_{fc}-\frac{1}{8X^{3}}\varepsilon^{abcd}R_{cmef}\phi_{a}\phi^{n}\phi^{m}\phi^{e}\phi_{d}^{f}\phi_{bn}\nonumber \\
	& \equiv & -F_{6}^{(1;2,0)}-F_{8}^{(1;2,0)}, \label{d4cor1}
\end{eqnarray}
\begin{eqnarray}
	\varepsilon_{ijk}K^{im}K^{jn}\nabla_{m}K_{n}^{k} & \rightarrow & \varepsilon^{abcd}u_{a}K_{b}^{e}K_{c}^{f}\mathrm{D}_{f}K_{de}\nonumber \\
	& = & -\frac{1}{8X^{3}}\varepsilon^{abcd}\phi_{a}\phi^{e}\left(-2XR_{efcm}\phi_{b}^{f}\phi_{d}^{m}+R_{efcm}\phi^{m}\phi^{n}\phi_{nb}\phi_{d}^{f}\right)\nonumber \\
	& \equiv & F_{5}^{(1;2,0)}-F_{8}^{(1;2,0)},
\end{eqnarray}
\begin{eqnarray}
	\varepsilon_{ijk}K^{mn}K_{m}^{i}\nabla^{j}K_{n}^{k} & \rightarrow & \varepsilon^{abcd}u_{a}K_{b}^{f}K_{f}^{e}\mathrm{D}_{c}K_{de}\nonumber \\
	& = & -\frac{1}{16X^{3}}\varepsilon^{abcd}\phi_{a}\phi^{e}\left(2XR_{efcd}\phi_{b}^{m}\phi_{m}^{f}+R_{cdef}\phi^{n}\phi^{m}\phi_{m}^{f}\phi_{nb}\right)\nonumber \\
	& \equiv & -\frac{1}{2}\left(F_{2}^{(1;2,0)}+F_{7}^{(1;2,0)}\right),
\end{eqnarray}
\begin{eqnarray}
	\varepsilon_{ijk}K_{l}^{i}a^{j}\nabla^{k}a^{l} & \rightarrow & \varepsilon^{abcd}u_{a}K_{r}^{b}a^{c}\mathrm{D}^{d}a^{r}\nonumber \\
	& = & -\frac{1}{8X^{3}}\varepsilon^{abcd}\phi_{a}\phi^{e}\phi^{f}\phi_{eb}\phi_{c}^{r}\phi_{drf}\nonumber \\
	& \equiv & -F_{6}^{(0;2,1)}=\frac{1}{2}F_{7}^{(1;2,0)}-\nabla_{d}\left(\frac{1}{\sigma^{6}}\varepsilon_{abc}^{\phantom{abc}d}\phi^{e}\phi^{f}\phi^{a}\phi_{e}^{b}\phi^{cm}\phi_{mf}\right),
\end{eqnarray}
where we have used (\ref{app:F0216red}),
\begin{eqnarray}
	\varepsilon_{ijk}R_{l}^{i}\nabla^{j}K^{kl} & \rightarrow & \varepsilon^{abcd}u_{a}\,{}^{3}\!R_{b}^{e}\mathrm{D}_{c}K_{de}\nonumber \\
	& = & \frac{1}{16X^{3}}\varepsilon^{abcd}\Big\{-4X^{2}R_{a}^{e}R_{cdef}\phi^{f}\phi_{b}-2X\phi_{b}\phi^{m}\phi^{f}\phi^{n}R_{am\phantom{e}n}^{\phantom{am}e}R_{cdef}\nonumber \\
	&  & \hspace{4em}+2X\phi_{a}\phi^{e}\left(R_{cdef}\phi_{b}^{f}\phi_{r}^{r}-R_{cdef}\phi_{b}^{m}\phi_{m}^{f}\right)\nonumber \\
	&  & \hspace{4em}+R_{cdef}\phi_{a}\phi^{e}\phi^{n}\phi^{m}\left(-\phi_{m}^{f}\phi_{nb}+\phi_{b}^{f}\phi_{mn}\right)\Big\}\nonumber \\
	& \equiv & -\frac{1}{2}\left(F_{4}^{(2;0,0)}+F_{5}^{(2;0,0)}+F_{2}^{(1;2,0)}+F_{7}^{(1;2,0)}-F_{9}^{(1;2,0)}-F_{10}^{(1;2,0)}\right),\qquad 
\end{eqnarray}
and
\begin{eqnarray}
	\varepsilon_{ijk}KK_{l}^{i}\nabla^{j}K^{kl} & \rightarrow & \varepsilon^{abcd}u_{a}KK_{b}^{e}\mathrm{D}_{c}K_{de}\nonumber \\
	& = & -\frac{1}{16X^{3}}\varepsilon^{abcd}\phi_{a}\phi^{e}\left(2R_{cdef}X\phi_{b}^{f}\phi_{r}^{r}+R_{cdrs}\phi^{r}\phi^{f}\phi_{b}^{s}\phi_{ef}\right)\nonumber \\
	& \equiv & -\frac{1}{2}\left(F_{9}^{(1;2,0)}+F_{10}^{(1;2,0)}\right). \label{d4cor6}
\end{eqnarray}

The last 2 monomials in the complete basis (\ref{Z4}) involve $F$ or $\pounds_{\bm{u}}K_{ij}$, which were not considered previously.
Their covariant correspondences are
\begin{eqnarray}
	\varepsilon_{ijk}FK_{l}^{i}\nabla^{j}K^{kl} & \rightarrow & \varepsilon^{abcd}u_{a}FK_{fb}\mathrm{D}_{c}K_{d}^{f}\nonumber \\
	& = & -\frac{1}{16X^{3}}\varepsilon^{abcd}R_{cdrs}\phi_{a}\phi^{r}\phi_{b}^{s}\phi_{fe}\phi^{e}\phi^{f}\nonumber \\
	& \equiv & -\frac{1}{2}F_{10}^{(1;2,0)}, \label{d4cor7}
\end{eqnarray}
and
\begin{eqnarray}
	\varepsilon_{ijk}a^{i}K^{jl}\pounds_{\bm{u}}K_{l}^{k} & \rightarrow & \varepsilon^{abcd}u_{a}a_{b}K_{c}^{f}h_{d}^{e}\pounds_{\bm{u}}K_{ef}\nonumber \\
	& = & \frac{1}{8X^{3}}\varepsilon^{abcd}\phi_{a}\phi^{e}\phi^{f}\phi_{eb}\left(\phi_{drf}\phi_{c}^{r}+R_{dfrs}\phi^{r}\phi_{c}^{s}\right)\nonumber \\
	& \equiv & F_{6}^{(0;2,1)}-F_{8}^{(1;2,0)}\nonumber \\
	& \equiv & -\frac{1}{2}F_{7}^{(1;2,0)}-F_{8}^{(1;2,0)}+\nabla_{d}\left(\frac{1}{\sigma^{6}}\varepsilon_{abc}^{\phantom{abc}d}\phi^{e}\phi^{f}\phi^{a}\phi_{e}^{b}\phi^{cm}\phi_{mf}\right), \label{d4cor8}
\end{eqnarray}
where again we have used (\ref{app:F0216red}).
We emphasize that the $[FK\nabla K]$ term and $[aK\pounds K]$ are independent in the sense that they cannot be reduced or related to each other by integration by parts.

To summarize, we find 7 combinations of GST monomials, given in (\ref{d3cor}) and (\ref{d4cor1})-(\ref{d4cor6}), which are free of the Ostrogradsky ghost in the unitary gauge.
On the other hand, the 2 combinations (\ref{d4cor7}) and (\ref{d4cor8}) contain linear terms in either $\pounds_{\bm{u}}K_{ij}$ or $F$, which are thus potentially risky even in the unitary gauge.

It is convenient to further combine (\ref{d3cor})-(\ref{d4cor6}) to obtain a set of 7 independent, ghost-free parity-violating Lagrangians.
For $d=3$, we choose 
	\begin{equation}
		\mathcal{L}_{1}\equiv F_{1}^{(1;1,0)}=\frac{1}{\sigma^{3}}\varepsilon_{abcd}R_{ef}^{\phantom{ef}cd}\phi^{a}\phi^{e}\phi^{bf},\label{Lag1}
	\end{equation}
which is, in fact, the single GST monomial of the $(1;1,0)$ category, i.e., linear in both the curvature tensor and the second-order derivative of the scalar field.
For $d=4$, there are 5 combinations of the $(1;2,0)$ category, i.e., linear in the curvature tensor and quadratic in the second derivative of the scalar field, which we choose to be
	\begin{eqnarray}
		\mathcal{L}_{2} & \equiv & F_{2}^{(1;2,0)}=\frac{1}{\sigma^{4}}\varepsilon_{abcd}R_{ef}^{\phantom{ef}cd}\phi^{a}\phi^{e}\phi_{m}^{b}\phi^{fm},
	\end{eqnarray}
	\begin{equation}
		\mathcal{L}_{3}\equiv F_{7}^{(1;2,0)}=\frac{1}{\sigma^{6}}\varepsilon_{abcd}R_{ef}^{\phantom{ef}cd}\phi^{m}\phi^{n}\phi^{e}\phi^{a}\phi_{m}^{f}\phi_{n}^{b},
	\end{equation}
	\begin{eqnarray}
		\mathcal{L}_{4} & \equiv & F_{6}^{(1;2,0)}+F_{8}^{(1;2,0)}\nonumber \\
		& = & \frac{1}{\sigma^{4}}\varepsilon_{abcd}R^{amen}\phi^{b}\phi^{f}\phi_{e}^{c}\phi_{f}^{d}\left(g_{mn}+\frac{1}{2X}\phi_{n}\phi_{m}\right),
	\end{eqnarray}
	\begin{eqnarray}
		\mathcal{L}_{5} & \equiv & F_{5}^{(1;2,0)}-F_{8}^{(1;2,0)}\nonumber \\
		& = & \frac{1}{\sigma^{4}}\varepsilon_{abcd}R_{ef}^{\phantom{ef}cm}\phi^{a}\phi^{e}\phi^{bf}\phi^{dn}\left(g_{mn}+\frac{1}{2X}\phi_{m}\phi_{n}\right),
	\end{eqnarray}
	\begin{eqnarray}
		\mathcal{L}_{6} & \equiv & F_{9}^{(1;2,0)}+F_{10}^{(1;2,0)}\nonumber \\
		& = & \frac{1}{\sigma^{4}}\varepsilon_{abcd}R_{ef}^{\phantom{ef}cd}\phi^{a}\phi^{e}\phi^{bf}\phi^{mn}\left(g_{mn}+\frac{1}{2X}\phi_{m}\phi_{n}\right).
	\end{eqnarray}
There is also a combination of the $(2;0,0)$ category, i.e., quadratic in the curvature tensor and without higher order derivatives of the scalar field, which we choose to be
	\begin{eqnarray}
		\mathcal{L}_{7} & \equiv & F_{4}^{(2;0,0)}+F_{5}^{(2;0,0)}\nonumber \\
		& = & \frac{1}{\sigma^{2}}\varepsilon_{abcd}R_{ef}^{\phantom{ef}cd}R^{amen}\phi^{b}\phi^{f}\left(g_{mn}+\frac{1}{2X}\phi_{m}\phi_{n}\right). \label{Lag7}
	\end{eqnarray}
The 7 Lagrangians (\ref{Lag1})-(\ref{Lag7}) are the main results in this work. Due to their generality and importance, we dub them the ``Qi-Xiu'' Lagrangians for the sake of brevity\footnote{``Qi-Xiu'' stands for ``Seven Constellations'' in Classical Chinese.}.

Recall that we have in total 17 GST monomials of $d=4$ (see Table \ref{tab:gstmono}). According to (\ref{app:F0216red}) and (\ref{app:F1014red}), $F^{(0;2,1)}_{6}$ and $F^{(1;0,1)}_{4}$ can be reduced by integrations by parts. Moreover, there are 7 identities in the unitary gauge among the rest 15 GST monomials of categories $(1;2,0)$ and $(2;0,0)$, which are shown in (\ref{app:ugid1})-(\ref{app:ugid7}). As a result, besides the 2 potentially dangerous terms, we are left with exactly 7 independent GST terms that are ghost-free in the unitary gauge. We thus conclude that up to $d=4$, any parity-violating scalar-tensor theory that is ghost-free in the unitary gauge can be expressed as
	\begin{equation}
		\mathcal{L} = \sum_{a=1}^{7} C_{a}(\phi,X)\mathcal{L}_{a},
	\end{equation}
that is, linear combinations of the ``Qi-Xiu'' Lagrangians with coefficients $C_{a}$ being general functions of $\phi$ and $X$.

\subsection{Comparing with the existing theories} \label{subsec:comp}

According to the previous section, up to $d=4$, there are, in total, 9 independent SCG monomials, of which 7 contain no $F$ or $\pounds_{\bm{u}}K_{ij}$ terms and thus are automatically healthy, and 2 are linear in $F$ or $\pounds_{\bm{u}}K_{ij}$ and thus are dangerous. Their generally covariant correspondence is 9 GST polynomials.
Any parity-violating GST polynomial up to $d=4$ can be expressed as a linear combination of these 9 combinations up to the identities in the unitary gauge (\ref{app:ugid1})-(\ref{app:ugid7}).

Let us take Chern-Simons gravity as an example.
The usual Chern-Simons term corresponds to the monomial $F_{1}^{(2;0,0)}$ in our classification of GST monomials.
The ``covariant'' 3+1 decomposition of Chern-Simons gravity with coefficient $f(\phi,X)$ is given by
	\begin{eqnarray}
	\mathcal{L}_{\mathrm{CS}} & = &f(\phi,X)F_{1}^{(2;0,0)} \nonumber\\ 
	& = & f(\phi,X)n^{a}\varepsilon_{abcd}\Big(-2\mathrm{D}^{e}a^{b}\mathrm{D}^{c}K_{e}^{d}+2\mathrm{D}^{e}K^{fb}K_{e}^{c}K_{f}^{d}-2\mathrm{D}^{b}K^{fc}K_{f}^{p}K_{p}^{d} \nonumber \\ 
	& &\qquad\qquad\qquad +{}^{3}\!R_{ef}^{\phantom{ef} cd}\mathrm{D}^{e}K^{fb}+2h^{pd}\mathrm{D}^{b}K^{fc}\pounds_{\bm{n}}K_{fp}\Big), \label{CSug}
	\end{eqnarray}
where the covariant 3+1 decomposition is performed with respect to an arbitrary normal vector $n^{a}$ (with no relation to $\phi$), i.e., no specific gauge is taken.
Ostrogradsky ghosts would appear in this case since the second-order time derivative of $h_{ij}$, namely $\pounds_{\bm{n}}K_{ij}$, is kinetically mixed with the dynamical scalar field through the function of $X$, which cannot be reduced by integration by parts.
Even in the unitary gauge with $n_{a} = u_{a} \equiv -N \nabla_{a} \phi$, the risky term $\pounds_{\bm{u}}K_{ij}$ will still be present, although it only appears linearly in the Lagrangian.
This is consistent with the previous analysis since $F_{1}^{(2;0,0)}$ is not ghost-free by itself.

As a special case, if the coefficient $f$ is a function of $\phi$ only, the term linear in $\pounds_{\bm{n}}K_{ij}$ can be reduced by integration by parts, which yields \cite{Gao:2019liu}
\begin{eqnarray}
f(\phi)F_{1}^{(2;0,0)} & \xlongequal{\mathrm{u.g.}} & 8\varepsilon^{ijk}f\Big(K_{il}K^{lm}\nabla_{j}K_{km}+K_{i}^{l}K_{j}^{m}\nabla_{m}K_{kl}-KK_{i}^{l}\nabla_{j}K_{kl}\\
&  & \qquad-2\,{}^{3}\!R_{i}^{l}\nabla_{j}K_{kl}-\frac{1}{N}\frac{\dot{f}}{f}K_{i}^{l}\nabla_{j}K_{kl}-\frac{2}{N}\nabla_{i}K_{jl}\nabla_{k}\nabla^{l}N\Big), \label{CS_fP_ug}
\end{eqnarray}
in the unitary gauge after fixing the spatial coordinates.
It is transparent that in the unitary gauge, Chern-Simons gravity with coefficient $f(\phi)$ reduces to the form of a ghost-free SCG (i.e., with 3 DoFs).
In particular, it takes the form of a linear combination of 6 monomials in (\ref{Z3}) and (\ref{Z4}).

Similar analysis can be performed for chiral scalar-tensor theories proposed in \cite{Crisostomi:2017ugk}, in which three classes of Lagrangians without the Ostrogradsky ghosts in the unitary gauge were identified.
The first class of Lagrangian is a linear combination of the following four terms:
\begin{equation}
	\mathcal{L}_{\text{PV}1}	=
	\sum_{n=1}^{4}a_{n}(\phi,X)\mathcal{A}_{n}, 
\end{equation}
where $\mathcal{A}_{1},\cdots,\mathcal{A}_{4}$ correspond to $(2;0,0)$ category in our notation\footnote{Here the correspondences between $\mathcal{A}_{1},\cdots,\mathcal{A}_{4}$ and $L_{1},\cdots,L_{4}$ in \cite{Crisostomi:2017ugk} are $\mathcal{A}_{1}=L_{1}$, $\mathcal{A}_{2}=L_{3}$, $\mathcal{A}_{3}=L_{2}$ and $\mathcal{A}_{4}=L_{4}$. Note there are in total 5 linearly independent monomials in $(2;,0,0)$ category (see (\ref{app:F2001})-(\ref{app:F2005})). $F^{(2;0,0)}_{5}$ is not considered in \cite{Crisostomi:2017ugk}.},
	\begin{eqnarray}
		\mathcal{A}_{1} & = & \varepsilon^{abcd}R_{cdef}R_{ab\phantom{e}g}^{\phantom{ab}e}\phi^{f}\phi^{g}\equiv-\sigma^{2}F_{2}^{(2;0,0)}, \label{calA1}\\
		\mathcal{A}_{2} & = & \varepsilon^{abcd}R_{cdef}R_{\phantom{f}b}^{f}\phi^{e}\phi_{a}\equiv\sigma^{2}F_{4}^{(2;0,0)},\\
		\mathcal{A}_{3} & = & \varepsilon^{abcd}R_{cdef}R_{ag}^{\phantom{ag}ef}\phi_{b}\phi^{g}\equiv\sigma^{2}F_{3}^{(2;0,0)},\\
		\mathcal{A}_{4} & = & \varepsilon^{abcd}R_{cdef}R_{\phantom{ef}ab}^{ef}\phi_{g}\phi^{g}\equiv-\sigma^{2}F_{1}^{(2;0,0)}, \label{calA4}
	\end{eqnarray}
where $a_{1},\cdots,a_{4}$ are general functions of $\phi$ and $2X\equiv -(\partial\phi)^2 \equiv \sigma^2$. In the above we have explicitly denoted these terms in the notation of GST monomials $F^{(2;,0,0)}_{n}$. 
According to (\ref{F2001dec})-(\ref{F2005dec}), in the unitary gauge, there are terms linear in $\pounds_{\bm{u}} K_{ij}$, which may be dangerous due to the second order time derivative encoded in $\pounds_{\bm{u}}K_{ij}$.
The coefficient of the combination of such terms is proportional to $4a_{1}+2a_{2}+a_{3}+8a_{4}$. Therefore, the degenerate condition
	\begin{equation}
		4a_{1}+a_{2}+2a_{3}+8a_{4}=0, \label{pvmod1cond}
	\end{equation}
identified in \cite{Crisostomi:2017ugk} can be regarded as requiring the vanishing of the dangerous term $[\pounds K\nabla K]$ in the unitary gauge.

Since the 4 coefficients $a_{1},\cdots, a_{4}$ are subject to a single constraint (\ref{pvmod1cond}), we have 3 combinations of terms:
	\begin{eqnarray}
		\mathcal{O}_{1} & \equiv & \mathcal{A}_{1}-\frac{1}{2}\mathcal{A}_{4}=\varepsilon^{abcd}R_{cdef}\left(R_{ab\phantom{e}g}^{\phantom{ab}e}\phi^{f}\phi^{g}-\frac{1}{2}R_{\phantom{ef}ab}^{ef}\phi_{g}\phi^{g}\right), \label{pv1O1}\\
		\mathcal{O}_{2} & \equiv & \mathcal{A}_{2}-\frac{1}{8}\mathcal{A}_{4}=\varepsilon^{abcd}R_{cdef}\left(R_{\phantom{f}b}^{f}\phi^{e}\phi_{a}-\frac{1}{8}R_{\phantom{ef}ab}^{ef}\phi_{g}\phi^{g}\right), \label{pv1O2}\\
		\mathcal{O}_{3} & \equiv & \mathcal{A}_{3}-\frac{1}{4}\mathcal{A}_{4}=\varepsilon^{abcd}R_{cdef}\left(R_{ag}^{\phantom{ag}ef}\phi_{b}\phi^{g}-\frac{1}{4}R_{\phantom{ef}ab}^{ef}\phi_{g}\phi^{g}\right), \label{LPVc1}
	\end{eqnarray}
which are free of the Ostrogradsky ghost in the unitary gauge. Indeed, in the unitary gauge one finds that \cite{Gao:2019liu}
	\begin{equation}
		\mathcal{O}_{1}^{\mathrm{(u.g.)}}=-\frac{8}{N^{2}}\varepsilon_{ijk}K^{li}K^{mj}\nabla_{m}K_{l}^{\phantom{l}k}+\frac{8}{N^{2}}\varepsilon_{ijk}\,{}^{3}\!R^{li}\nabla^{k}K_{l}^{\phantom{l}j},
	\end{equation}
	\begin{eqnarray}
		\mathcal{O}_{2}^{\mathrm{(u.g.)}} & = & -\frac{2}{N^{2}}\varepsilon_{ijk}K^{li}K^{mj}\nabla_{m}K_{l}^{\phantom{l}k}-\frac{2}{N^{2}}\varepsilon_{ijk}\left(K_{m}^{\phantom{l}i}K^{lm}-KK^{li}\right)\nabla^{k}K_{l}^{\phantom{l}j}\nonumber \\
		&  & +\frac{4}{N^{2}}\varepsilon_{ijk}\,{}^{3}\!R^{li}\nabla^{k}K_{l}^{\phantom{l}j},
	\end{eqnarray}
and $\mathcal{O}_{3}^{\mathrm{(u.g.)}} \equiv  0$\footnote{Note $\mathcal{L}_{3}$ in (\ref{LPVc1}) is nothing but eq. (3.5) in \cite{Crisostomi:2017ugk}, which has been pointed out to be vanishing in the unitary gauge.}.
Therefore, when restricted to the 4 GST monomials in (\ref{calA1})-(\ref{calA4}), there are 2 independent ghost-free combinations, given in (\ref{pv1O1}) and (\ref{pv1O2}).

The second class of Lagrangian is composed of a single term belonging to the $(1;1,0)$ category,
	\begin{equation}
		\mathcal{L}_{\text{PV}2} = b(\phi,X)\mathcal{B}, \qquad \text{with}\quad \mathcal{B}=\varepsilon^{abcd}R_{cdef}\phi_{a}\phi^{e}\phi_{b}^{f}\equiv\sigma^{3}F_{1}^{(1;1,0)},
	\end{equation}
which is healthy in the unitary gauge since
	\begin{equation}
		\mathcal{B}^{\mathrm{(u.g.)}} = \frac{2}{N^{3}}\varepsilon_{ijk}K^{li}\nabla^{j}K_{l}^{\phantom{l}k}.
	\end{equation}

The third class of Lagrangian is the linear combination of 6 monomials belonging to the $(1;2,0)$ category,
\begin{equation}
	\mathcal{L}_{\text{PV}3}	=
	\sum_{n=1}^{6}c_{n}(\phi,X) \mathcal{C}_{n}, 
\end{equation}
with\footnote{Here the correspondences between $\mathcal{C}_{1},\cdots,\mathcal{C}_{6}$
	and $L_{2},\cdots,L_{7}$ in eq. (3.13) of \cite{Crisostomi:2017ugk}
	are $\mathcal{C}_{1}=-\sigma^{2}L_{2}$, $\mathcal{C}_{2}=L_{3}$,
	$\mathcal{C}_{3}=L_{5}$, $\mathcal{C}_{4}=L_{4}$, $\mathcal{C}_{5}=L_{6}$,
	$\mathcal{C}_{6}=L_{7}$. Recall that there are in total 10 linearly independent monomials in $(1;2,0)$ category (see (\ref{app:F1201})-(\ref{app:F12010})). Only 6 of them were considered in \cite{Crisostomi:2017ugk}.}
\begin{eqnarray}
	\mathcal{C}_{1} & = & \varepsilon^{abcd}R_{cdef}\phi_{a}^{e}\phi_{b}^{f}\phi_{g}\phi^{g}\equiv-\sigma^{4}F_{1}^{(1;2,0)}, \label{calC1}\\
	\mathcal{C}_{2} & = & \varepsilon^{abcd}R_{cdef}\phi_{a}^{e}\phi_{b}^{g}\phi^{f}\phi_{g}\equiv\sigma^{4}F_{3}^{(1;2,0)},\\
	\mathcal{C}_{3} & = & \varepsilon^{abcd}R_{cefg}\phi_{a}^{f}\phi_{b}^{g}\phi^{e}\phi_{d}\equiv-2\sigma^{4}F_{5}^{(1;2,0)},\\
	\mathcal{C}_{4} & = & \varepsilon^{abcd}R_{cdef}\phi_{a}^{e}\phi_{g}^{f}\phi_{b}\phi^{g}\equiv-\sigma^{4}F_{4}^{(1;2,0)},\\
	\mathcal{C}_{5} & = & \varepsilon^{abcd}R_{de}\phi_{a}^{e}\phi_{b}^{f}\phi_{c}\phi_{f}\equiv-\sigma^{4}F_{6}^{(1;2,0)},\\
	\mathcal{C}_{6} & = & \varepsilon^{abcd}R_{cdef}\phi^{e}\phi_{a}\phi_{b}^{f}\square\phi\equiv \sigma^{4}F_{9}^{(1;2,0)}, \label{calC6}
\end{eqnarray}
where $c_{1},\cdots,c_{6}$ are general functions of $\phi$ and $X$. 
After some manipulations, one can show that the coefficients of $\pounds_{\bm{u}}K_{ij}$ and $\pounds_{\bm{u}}N$ are proportional to $4c_{1}+2c_{2}+2c_{3}-c_{5}$ and $2c_{1}+c_{2}+c_{4}+c_{6}$, respectively.
In \cite{Crisostomi:2017ugk}, $c_{6}$ is set to be vanishing\footnote{This is to get rid of the dangerous term without fixing the unitary gauge.}.  Thus one requires
	\begin{eqnarray}
		4c_{1}+2c_{2}+2c_{3}-c_{5} & = & 0,\\
		2c_{1}+c_{2}+c_{4} & = & 0,
	\end{eqnarray}
in order to evade the Ostrogradsky ghost.
Now there are 5 coefficients $c_1,\cdots,c_5$ subject to 2 constraints. Therefore, we have 3 combinations of terms that are free of Ostrogradsky ghosts in the unitary gauge:
	\begin{eqnarray}
		\mathcal{O}_{1} & \equiv & \mathcal{C}_{1}-2\mathcal{C}_{4}+4\mathcal{C}_{5}\nonumber \\
		& = & \varepsilon^{abcd}\left[R_{cdef}\left(\phi_{b}^{f}\phi_{g}\phi^{g}-2\phi_{g}^{f}\phi_{b}\phi^{g}\right)+4R_{de}\phi_{b}^{f}\phi_{c}\phi_{f}\right]\phi_{a}^{e}, \label{pv3O1}
	\end{eqnarray}
	\begin{eqnarray}
		\mathcal{O}_{2} & \equiv & \mathcal{C}_{2}-\mathcal{C}_{4}+2\mathcal{C}_{5}\nonumber \\
		& = & \varepsilon^{abcd}\left[R_{cdef}\left(\phi_{b}^{g}\phi^{f}\phi_{g}-\phi_{g}^{f}\phi_{b}\phi^{g}\right)+2R_{de}\phi_{b}^{f}\phi_{c}\phi_{f}\right]\phi_{a}^{e},
	\end{eqnarray}
and
	\begin{equation}
		\mathcal{O}_{3}\equiv\mathcal{C}_{3}+2\mathcal{C}_{5}=\varepsilon^{abcd}\left(R_{cgef}\phi^{g}\phi_{d}+2R_{de}\phi_{c}\phi_{f}\right)\phi_{a}^{e}\phi_{b}^{f}.
	\end{equation}
This can be checked explicitly. In the unitary gauge \cite{Gao:2019liu},
	\begin{eqnarray}
		\mathcal{O}_{1}^{\mathrm{(u.g.)}} & = & \frac{4}{N^{4}}\varepsilon_{ijk}\left(K^{li}K^{mj}\nabla_{m}K_{l}^{\phantom{b}k}-\frac{1}{N}K^{lj}\,{}^{3}\!R_{l}^{\phantom{l}k}\nabla^{i}N\right),
	\end{eqnarray}
while $\mathcal{O}_{2}^{\mathrm{(u.g.)}}=0$ and $\mathcal{O}_{3}^{\mathrm{(u.g.)}}=\frac{1}{2}\mathcal{L}_{1}^{\mathrm{(u.g.)}}$.
Therefore, when restricted to the 6 GST monomials in (\ref{calC1})-(\ref{calC6}), there is only one independent ghost-free combination given in (\ref{pv3O1}).

\section{Conclusions} \label{sec:con}

Recently, there has been an increasing interest in studying gravitational theories with parity violation.
In this work, we have investigated the scalar-tensor theory with parity violation. In particular, we are looking for the special combinations of the scalar-tensor monomials that are free of the Ostrogradsky ghost in the unitary gauge, i.e., when the scalar field possesses a timelike gradient.

Since the generally covariant scalar-tensor theory (GST) in the unitary gauge takes the form of spatially covariant gravity (SCG), in Sec. \ref{sec:scg}, we describe the general framework of SCG theory and the classification of SCG monomials.
We extend the SCG theory by introducing Lie derivatives of the lapse function $F\equiv \pounds_{\bm{u}} \ln N$ and the extrinsic curvature $\pounds_{\bm{u}}K_{ij}$. This is not only because they are the natural building blocks of SCG, but also because they necessarily arise in the decomposition of scalar-tensor theory in the unitary gauge up to $d=4$ with $d$ the total number of derivatives.

In Sec. \ref{sec:pvscg}, by including $F$ and $\pounds_{\bm{u}}K_{ij}$ as the building blocks of SCG, we exhausted the SCG monomials up to $d=4$, which are classified and summarized in Table \ref{SCG_pv_mono}.
Based on this classification, we obtain the complete basis for the parity-violating SCG polynomials of $d=3$ and $d=4$, which are given in (\ref{Z3}) and (\ref{Z4}), respectively.
In total, there are 9 independent SCG monomials with parity violation, of which 7 contain no higher temporal derivatives and are thus automatically free of ghosts, while 2 involve Lie derivatives of the extrinsic curvature and the lapse function and are thus potentially risky. Our analysis thus generalizes the previous result presented in \cite{Gao:2020yzr}.

Our final goal is to find the ghost-free combinations of GST monomials, which is performed in Sec. \ref{sec:gfgst}.
To this end, in Sec. \ref{subsec:dec} we derive the decomposition of the GST monomials in the unitary gauge.
The resulting expressions show the dangerous temporal derivative terms explicitly and thus give us guidance on finding the ghost-free combinations.
The main results in this work are presented in Sec. \ref{subsec:gfst}, where we derive the generally covariant correspondence of the 9 parity-violating SCG monomials in the complete basis (\ref{Z3}) and (\ref{Z4}). Since 7 out of the 9 SCG monomials are ghost-free, there must be 7 scalar-tensor Lagrangians that are ghost-free in the unitary gauge, which we choose to be (\ref{Lag1})-(\ref{Lag7}) and dub them the ``Qi-Xiu'' Lagrangians for short.
Up to $d=4$, since any parity-violating GST polynomial in the unitary gauge takes the form of a linear combination of the 9 parity-violating SCG monomials in the complete basis, we conclude that, up to 7 identities in the unitary gauge (\ref{app:ugid1})-(\ref{app:ugid7}), the ``Qi-Xiu'' Lagrangians (\ref{Lag1})-(\ref{Lag7}) are the most general parity-violating scalar-tensor theories that are ghost-free in the unitary gauge up to $d=4$.
As shown in Sec. \ref{subsec:comp}, our results have included the Chern-Simons term as well as the chiral scalar-tensor theory proposed in \cite{Crisostomi:2017ugk} as special cases.

\acknowledgments

We thank Yi-Fu Cai, Yang Yu and Zhiyu Lu for valuable discussions. 
This work was partly supported by the National Natural Science Foundation of China (NSFC) under the grants  No. 12347137 (YMH) and No. 11975020 (XG).

\appendix

\section{Reduction of $[\pounds K \nabla K]$ term} \label{app:rel}

In this appendix, we will show that $\varepsilon_{ijk}\pounds_{\bm{u}}K_{l}^{i}\nabla^{j}K^{kl}$ can be reduced by integration by parts. When performing integration by parts regarding the SCG terms, it is more convenient to use the generally covariant corresponding expressions. In our case, it is given by
	\begin{equation}
		\varepsilon^{ijk}\pounds_{\bm{u}}K_{li}\nabla_{j}K_{k}^{l}\rightarrow f\varepsilon^{\bm{u}\hat{b}\hat{c}\hat{d}}h^{ef}\pounds_{\bm{u}}K_{be}\mathrm{D}_{c}K_{df} \equiv \mathcal{O},
	\end{equation}
where we define $\varepsilon^{\bm{u}\hat{b}\hat{c}\hat{d}}\equiv u_{a}h_{b'}^{b}h_{c'}^{c}h_{d'}^{d}\varepsilon^{ab'c'd'}$ for later convenience.

There are two equivalent approaches to performing the integrations by parts: either using the Lie derivative and intrinsic derivative directly, or using the 4-dimensional covariant derivative $\nabla_{a}$ by expanding Lie/intrinsic derivatives in terms of the covariant derivative. Here we choose the former. We have
\begin{eqnarray}
	\mathcal{O} & = & f\varepsilon^{\bm{u}\hat{b}\hat{c}\hat{d}}h^{ef}\pounds_{\bm{u}}K_{be}\mathrm{D}_{c}K_{df}\nonumber \\
	& = & \pounds_{\bm{u}}\left(f\varepsilon^{\bm{u}\hat{b}\hat{c}\hat{d}}h^{ef}K_{be}\mathrm{D}_{c}K_{df}\right)-K_{be}\pounds_{\bm{u}}\left(f\varepsilon^{\bm{u}\hat{b}\hat{c}\hat{d}}h^{ef}\mathrm{D}_{c}K_{df}\right)\nonumber \\
	& \simeq & -Kf\varepsilon^{\bm{u}\hat{b}\hat{c}\hat{d}}K_{b}^{f}\mathrm{D}_{c}K_{df}-K_{be}\pounds_{\bm{u}}\left(f\varepsilon^{\bm{u}\hat{b}\hat{c}\hat{d}}h^{ef}\right)\mathrm{D}_{c}K_{df}-K_{be}f\varepsilon^{\bm{u}\hat{b}\hat{c}\hat{d}}h^{ef}\pounds_{\bm{u}}\mathrm{D}_{c}K_{df}\nonumber \\
	& = & -f\varepsilon^{\bm{u}\hat{b}\hat{c}\hat{d}}KK_{b}^{f}\mathrm{D}_{c}K_{df}-K_{be}\pounds_{\bm{u}}\left(f\varepsilon^{\bm{u}\hat{b}\hat{c}\hat{d}}h^{ef}\right)\mathrm{D}_{c}K_{df}-K_{b}^{f}f\varepsilon^{\bm{u}\hat{b}\hat{c}\hat{d}}\left[\pounds_{\bm{u}},\mathrm{D}_{c}\right]K_{df}\nonumber \\
	&  & -\mathrm{D}_{c}\left(K_{b}^{f}f\varepsilon^{\bm{u}\hat{b}\hat{c}\hat{d}}\pounds_{\bm{u}}K_{df}\right)+\mathrm{D}_{c}\left(f\varepsilon^{\bm{u}\hat{b}\hat{c}\hat{d}}\right)K_{b}^{f}\pounds_{\bm{u}}K_{df}+\mathrm{D}_{c}K_{b}^{f}f\varepsilon^{\bm{u}\hat{b}\hat{c}\hat{d}}\pounds_{\bm{u}}K_{df},\label{calOint}
\end{eqnarray}
where we have used that for a scalar field $\Phi$, $\pounds_{\bm{u}}\Phi=u^{a}\nabla_{a}\Phi\simeq-K\Phi$,
and the commutator is defined by
\begin{equation}
	\left[\pounds_{\bm{u}},\mathrm{D}_{c}\right]K_{df}=\pounds_{\bm{u}}\left(\mathrm{D}_{c}K_{df}\right)-\mathrm{D}_{c}\left(\pounds_{\bm{u}}K_{df}\right).
\end{equation}

For the second term in (\ref{calOint}), by using
\[
\pounds_{\bm{u}}f=\frac{\partial f}{\partial\phi}\pounds_{\bm{u}}\phi+N\frac{\partial f}{\partial N}F,
\]
with $F\equiv\frac{1}{N}\pounds_{\bm{u}}N$,
\[
\pounds_{\bm{u}}h^{ab}=-2K^{ab}+2u^{(a}a^{b)},
\]
and
\[
\pounds_{\bm{u}}\varepsilon^{\bm{n}\hat{a}\hat{b}\hat{c}}=a_{d}\varepsilon^{d\hat{a}\hat{b}\hat{c}}-\left(K_{e}^{a}-u^{a}a_{e}\right)\varepsilon^{\bm{u}e\hat{b}\hat{c}}-\left(K_{e}^{b}-u^{b}a_{e}\right)\varepsilon^{\bm{u}\hat{a}e\hat{c}}-\left(K_{e}^{c}-u^{c}a_{e}\right)\varepsilon^{\bm{u}\hat{a}\hat{b}e},
\]
we find
\begin{eqnarray}
	K_{be}\pounds_{\bm{u}}\left(f\varepsilon^{\bm{u}\hat{b}\hat{c}\hat{d}}h^{ef}\right)\mathrm{D}_{c}K_{df} & = & \left(\frac{\partial f}{\partial\phi}\pounds_{\bm{u}}\phi+N\frac{\partial f}{\partial N}F\right)\varepsilon^{\bm{u}\hat{b}\hat{c}\hat{d}}K_{b}^{f}\mathrm{D}_{c}K_{df}\nonumber \\
	&  & -3f\varepsilon^{\bm{u}\hat{b}\hat{c}\hat{d}}K_{m}^{f}K_{b}^{m}\mathrm{D}_{c}K_{df}-f\varepsilon^{\bm{u}\hat{b}\hat{c}\hat{d}}K_{b}^{f}K_{c}^{e}\mathrm{D}_{e}K_{df}.
\end{eqnarray}

For the third term in (\ref{calOint}), the commutator is given by
\[
\left[\pounds_{\bm{u}},\mathrm{D}_{c}\right]K_{df}=a_{c}\pounds_{\bm{u}}K_{df}+\Xi_{cdm}K_{\phantom{m}f}^{m}+\Xi_{cfm}K_{d}^{\phantom{d}m},
\]
with 
\[
\Xi_{cdm}=-\left(a_{c}K_{dm}+a_{d}K_{mc}-a_{m}K_{cd}\right)-\left(\mathrm{D}_{c}K_{dm}+\mathrm{D}_{d}K_{mc}-\mathrm{D}_{m}K_{cd}\right),
\]
and $\Xi_{cfm}$ is given accordingly. Thus
\begin{equation}
	K_{b}^{f}f\varepsilon^{\bm{u}\hat{b}\hat{c}\hat{d}}\left[\pounds_{\bm{u}},\mathrm{D}_{c}\right]K_{df}=-f\varepsilon^{\bm{u}\hat{b}\hat{c}\hat{d}}a_{b}K_{c}^{f}\pounds_{\bm{u}}K_{df}-2f\varepsilon^{\bm{u}\hat{b}\hat{c}\hat{d}}K_{b}^{f}K_{c}^{e}\mathrm{D}_{e}K_{df}.
\end{equation}

For the fourth term in (\ref{calOint}), we have
\begin{equation}
	\mathrm{D}_{c}\left(K_{b}^{f}f\varepsilon^{\bm{u}\hat{b}\hat{c}\hat{d}}\pounds_{\bm{u}}K_{df}\right)\simeq f\varepsilon^{\bm{u}\hat{b}\hat{c}\hat{d}}a_{b}K_{c}^{f}\pounds_{\bm{u}}K_{df},
\end{equation}
where we have used that for a tangent vector $A^{a}$, $\mathrm{D}_{a}A^{a}=\nabla_{a}A^{a}-a_{a}A^{a}$.

For the fifth term in (\ref{calOint}), we have
\begin{equation}
	\mathrm{D}_{c}\left(f\varepsilon^{\bm{u}\hat{b}\hat{c}\hat{d}}\right)K_{b}^{f}\pounds_{\bm{u}}K_{df}=-\varepsilon^{\bm{u}\hat{b}\hat{c}\hat{d}}N\frac{\partial f}{\partial N}a_{b}K_{c}^{f}\pounds_{\bm{u}}K_{df},
\end{equation}
where we have used $\mathrm{D}_{c}\varepsilon^{\bm{u}\hat{b}\hat{c}\hat{d}}=K_{ce}\varepsilon^{e\hat{b}\hat{c}\hat{d}}$.

Putting all the above together, and noting that the last term in (\ref{calOint})
is nothing but $-\mathcal{O}$, we get
\begin{eqnarray}
	\mathcal{O} & \simeq & -\frac{1}{2}f\varepsilon^{\bm{u}\hat{b}\hat{c}\hat{d}}KK_{b}^{f}\mathrm{D}_{c}K_{df}+\frac{3}{2}f\varepsilon^{\bm{n}\hat{b}\hat{c}\hat{d}}K_{m}^{f}K_{b}^{m}\mathrm{D}_{c}K_{df}+\frac{3}{2}f\varepsilon^{\bm{u}\hat{b}\hat{c}\hat{d}}K_{b}^{f}K_{c}^{e}\mathrm{D}_{e}K_{df}\nonumber \\
	&  & -\frac{1}{2}\left(\frac{\partial f}{\partial\phi}\pounds_{\bm{u}}\phi+N\frac{\partial f}{\partial N}F\right)\varepsilon^{\bm{u}\hat{b}\hat{c}\hat{d}}K_{b}^{f}\mathrm{D}_{c}K_{df}\nonumber \\
	&  & -\frac{1}{2}\varepsilon^{\bm{u}\hat{b}\hat{c}\hat{d}}N\frac{\partial f}{\partial N}a_{b}K_{c}^{f}\pounds_{\bm{u}}K_{df}.
\end{eqnarray}
In terms of spatial indices with $t=\phi$, 
\begin{eqnarray}
	\mathcal{O} & \simeq & -\frac{1}{2}f\varepsilon^{ijk}KK_{i}^{l}\nabla_{j}K_{kl}+\frac{3}{2}f\varepsilon^{ijk}K_{m}^{l}K_{i}^{m}\nabla_{j}K_{kl}+\frac{3}{2}f\varepsilon^{ijk}K_{i}^{l}K_{j}^{m}\nabla_{m}K_{kl}\nonumber \\
	&  & -\frac{1}{2}\left(\frac{\partial f}{\partial t}+N\frac{\partial f}{\partial N}F\right)\varepsilon^{ijk}K_{i}^{l}\nabla_{j}K_{kl}-\frac{1}{2}\varepsilon^{ijk}N\frac{\partial f}{\partial N}a_{i}K_{j}^{l}\pounds_{\bm{u}}K_{kl}.
\end{eqnarray}

To conclude, we have explicitly shown that $\varepsilon^{ijk}\pounds_{\bm{u}}K_{li}\nabla_{j}K_{k}^{l}$ is not independent, which can be reduced to a linear combination of $\left[KK\nabla K\right]$, $\left[K\nabla K\right]$, and $\left[aK\pounds K\right]$ terms by integrations by parts.

\section{Integrations by parts} \label{app:ibp}

For the $(1;2,0)$ category, the decomposition of the following GST monomials can be recast by integration by parts:
\begin{eqnarray}
	F_{1}^{(1;2,0)} & \xlongequal{\text{u.g.}} & 4\varepsilon_{abc}\Big(-\,{}^{3}\!R_{d}^{a}K^{db}a^{c}-2K_{d}^{a}a^{b}\mathrm{D}^{c}a^{d}-FK^{ad}\mathrm{D}^{b}K_{d}^{c}-K^{ad}K^{be}\mathrm{D}_{e}K_{d}^{c}\nonumber \\
	&  & \qquad-a^{a}K^{db}h^{ec}\pounds_{\boldsymbol{u}}K_{de}\Big)-4\nabla^{c}\left(\varepsilon_{abc}a^{a}a^{d}K_{d}^{b}\right), \label{F1201ibp}
\end{eqnarray}
\begin{equation}
	F_{2}^{(1;2,0)} \xlongequal{\text{u.g.}}  -2\varepsilon_{abc}\left(K_{e}^{a}a^{b}\mathrm{D}^{c}a^{e}+K_{d}^{a}K^{de}\mathrm{D}^{b}K_{e}^{c}\right)-2\nabla^{c}\left(\varepsilon_{abc}a^{a}a^{d}K_{d}^{b}\right), \label{F1202ibp}
\end{equation}
\begin{eqnarray}
	F_{3}^{(1;2,0)} & \xlongequal{\text{u.g.}} & 2\varepsilon_{abc}\left(2K_{d}^{a}a^{b}\mathrm{D}^{c}a^{d}+FK^{ad}\mathrm{D}^{b}K_{d}^{c}+a^{a}K^{db}h^{ec}\pounds_{\boldsymbol{u}}K_{de}\right)\nonumber \\
	&  & +2\nabla^{c}\left(\varepsilon_{abc}a^{a}a^{d}K_{d}^{b}\right), \label{F1203ibp}
\end{eqnarray}
\begin{equation}
	F_{4}^{(1;2,0)} \xlongequal{\text{u.g.}}  -2\varepsilon_{abc}\left(K_{e}^{a}a^{b}\mathrm{D}^{c}a^{e}+\,{}^{3}\!R_{d}^{a}K^{db}a^{c}+FK^{ad}\mathrm{D}^{b}K_{d}^{c}\right)-2\nabla^{c}\left(\varepsilon_{abc}a^{a}a^{d}K_{d}^{b}\right),\qquad \label{F1204ibp}
\end{equation}
and
\begin{equation}
	F_{7}^{(1;2,0)}\xlongequal{\text{u.g.}}2\varepsilon_{abc}K_{e}^{a}a^{b}\mathrm{D}^{c}a^{e}+2\nabla^{c}\left(\varepsilon_{abc}a^{a}a^{d}K_{d}^{b}\right). \label{F1207ibp}
\end{equation}

For the $(2;0,0)$ category, the decomposition of the following GST monomials can be  recast by integration by parts:
	\begin{eqnarray}
		F_{1}^{(2;0,0)} & \xlongequal{\text{u.g.}} & 8\varepsilon_{abc}\Big(2\,{}^{3}\!R^{ad}K_{d}^{b}a^{c}-2\,{}^{3}\!R^{da}\mathrm{D}^{b}K_{d}^{c}+K_{d}^{a}K^{de}\mathrm{D}^{b}K_{e}^{c}+K^{da}K^{eb}\mathrm{D}_{e}K_{d}^{c}\nonumber \\
		&  & \qquad-KK_{e}^{a}\mathrm{D}^{b}K^{ce}\Big)+8\mathcal{D},\label{F2001ibp}
	\end{eqnarray}
	\begin{eqnarray}
		F_{2}^{(2;0,0)} & \xlongequal{\text{u.g.}} & 4\varepsilon_{abc}\Big(2\,{}^{3}\!R^{ad}K_{d}^{b}a^{c}+K_{d}^{a}K^{de}\mathrm{D}^{b}K_{e}^{c}+3K^{da}K^{eb}\mathrm{D}_{e}K_{d}^{c}\nonumber \\
		&  & \qquad-KK_{e}^{a}\mathrm{D}^{b}K^{ce}\Big)+4\mathcal{D},\label{F2002ibp}
	\end{eqnarray}
	\begin{eqnarray}
		F_{3}^{(2;0,0)} & \xlongequal{\text{u.g.}} & 2\varepsilon_{abc}\Big(-2\,{}^{3}\!R^{ad}K_{d}^{b}a^{c}+2\,{}^{3}\!R^{da}\mathrm{D}^{b}K_{d}^{c}-K_{d}^{a}K^{de}\mathrm{D}^{b}K_{e}^{c}-K^{da}K^{eb}\mathrm{D}_{e}K_{d}^{c}\nonumber \\
		&  & \qquad+KK_{e}^{a}\mathrm{D}^{b}K^{ce}\Big)-2\mathcal{D},\label{F2003ibp}
	\end{eqnarray}
	\begin{eqnarray}
		F_{4}^{(2;0,0)} & \xlongequal{\text{u.g.}} & \varepsilon_{abc}\Big(-2\,{}^{3}\!R^{ad}K_{d}^{b}a^{c}-2\,{}^{3}\!R^{da}\mathrm{D}^{b}K_{d}^{c}+K_{d}^{a}K^{de}\mathrm{D}^{b}K_{e}^{c}-3K^{da}K^{eb}\mathrm{D}_{e}K_{d}^{c}\nonumber \\
		&  & \qquad-KK^{ea}\mathrm{D}^{c}K_{e}^{b}\Big)-\mathcal{D},\label{F2004ibp}
	\end{eqnarray}
and
	\begin{eqnarray}
		F_{5}^{(2;0,0)} & \xlongequal{\text{u.g.}} & \varepsilon_{abc}\Big(2\,{}^{3}\!R^{ad}K_{d}^{b}a^{c}+K_{d}^{a}K^{de}\mathrm{D}^{b}K_{e}^{c}+3K^{da}K^{eb}\mathrm{D}_{e}K_{d}^{c}\nonumber \\
		&  & \qquad-KK^{ea}\mathrm{D}^{b}K_{e}^{c}\Big)+\mathcal{D},\label{F2005ibp}
	\end{eqnarray}
where in (\ref{F2001ibp})-(\ref{F2005ibp}), $\mathcal{D}$ stands for the total derivative
	\begin{eqnarray}
		\mathcal{D} & = & 2\nabla^{c}\left(\varepsilon_{abc}\left(\mathrm{D}^{a}a^{d}+a^{a}a^{d}\right)K_{d}^{b}\right)\nonumber \\
		&  & +\nabla_{e}\left(n^{a}\varepsilon_{abcd}n^{e}K_{f}^{d}\nabla^{b}K^{cf}\right)+\nabla^{b}\left(n^{a}\varepsilon_{abcd}K^{cf}n^{e}\nabla_{e}K_{f}^{d}\right). \label{app:calD}
	\end{eqnarray}

In deriving the decomposition in the unitary gauge, we frequently make use of the following integrations by parts
\begin{eqnarray}
	\varepsilon_{abc}\mathrm{D}^{a}a^{d}\mathrm{D}^{c}K_{d}^{b} & = & -\varepsilon_{abc}\left(K_{e}^{a}a^{b}\mathrm{D}^{c}a^{e}-\,{}^{3}\!R_{e}^{a}K^{be}a^{c}\right)+\nabla^{c}\left(\varepsilon_{abc}\mathrm{D}^{a}a^{d}K_{d}^{b}\right),\\
	\varepsilon_{abc}a^{a}a^{d}\mathrm{D}^{c}K_{d}^{b} & = & \varepsilon_{abc}K_{e}^{a}a^{b}\mathrm{D}^{c}a^{e}+\nabla^{c}\left(\varepsilon_{abc}a^{a}a^{d}K_{d}^{b}\right),
\end{eqnarray}
and
\begin{eqnarray}
	\varepsilon^{abcd}u_{a}\pounds_{\bm{u}}K_{be}\mathrm{D}_{c}K_{d}^{e} & = & \frac{1}{2}\varepsilon_{abc}\left(3K_{e}^{a}K_{f}^{e}\mathrm{D}^{b}K^{cf}+3K^{ea}K^{rb}\mathrm{D}_{r}K_{e}^{c}-KK_{f}^{a}\mathrm{D}^{b}K^{cf}\right)\nonumber \\
	&  & +\frac{1}{2}\nabla_{e}\left(\varepsilon_{abc}u^{e}K_{f}^{c}\nabla^{a}K^{bf}\right)+\frac{1}{2}\nabla^{a}\left(\varepsilon_{abc}K^{bf}u^{e}\nabla_{e}K_{f}^{c}\right).
\end{eqnarray}
In the above, $\varepsilon_{abc}$ is defined in (\ref{eps3def}).
The purpose of extracting the total derivatives is to recast the expressions in terms of monomials in the SCG basis. One can also show that
\begin{eqnarray}
	\nabla^{c}\left(\varepsilon_{abc}a^{a}a^{d}K_{d}^{b}\right) & = & \varepsilon_{abc}\left(-K_{e}^{a}a^{b}\mathrm{D}^{c}a^{e}+a^{a}a^{d}\mathrm{D}^{c}K_{d}^{b}\right),\\
	\nabla^{c}\left(\varepsilon_{abc}\mathrm{D}^{a}a^{d}K_{d}^{b}\right) & = & \varepsilon_{abc}\left(K_{e}^{a}a^{b}\mathrm{D}^{c}a^{e}-\,{}^{3}\!R_{e}^{a}K^{be}a^{c}+\varepsilon_{abc}\mathrm{D}^{a}a^{d}\mathrm{D}^{c}K_{d}^{b}\right).
\end{eqnarray}

\section{Parity-violating GST monomials} \label{app:GSTmono}

The linearly independent GST monomials up to $d=4$ have been exhausted and classified in \cite{Gao:2020juc} (see also \cite{Gao:2020yzr}). In this appendix, we list the parity-violating monomials in the cases of $d=3$ and $d=4$ for convenience, which are summarized in Table \ref{tab:gstmono}.
\begin{center}
	\begin{table}[h]
		\renewcommand\arraystretch{1.6}
		\resizebox{\linewidth}{!}{
			\begin{tabular}{|c|c|c|c|c|}
				\hline 
				$d$ & Category & Unfactorizable & Factorizable & Number\tabularnewline
				\hline 
				3 & \{1;1,0\} & $F_{1}^{(1;1,0)}$ & $-$ & 1\tabularnewline
				&  &  &  & \tabularnewline
				\hline 
				4 & \{0;2,1\} & $F_{6}^{(0;2,1)}$ & $-$ & 1\tabularnewline
				\cline{2-5} \cline{3-5} \cline{4-5} \cline{5-5} 
				& \{1;2,0\} & $F_{1}^{(1;2,0)}$ &  & 10\tabularnewline
				&  & $F_{2}^{(1;2,0)},\quad F_{3}^{(1;2,0)},\quad F_{4}^{(1;2,0)},\quad F_{5}^{(1;2,0)},\quad F_{6}^{(1;2,0)}$ & $F_{9}^{(1;2,0)}=F_{1}^{(1;1,0)}E_{1}^{(0;1,0)}$ & \tabularnewline
				&  & $F_{7}^{(1;2,0)},\quad F_{8}^{(1;2,0)}$ & $F_{10}^{(1;2,0)}=F_{1}^{(1;1,0)}E_{2}^{(0;1,0)}$ & \tabularnewline
				\cline{2-5} \cline{3-5} \cline{4-5} \cline{5-5} 
				& \{2;0,0\} & $F_{1}^{(2;0,0)}$ & $-$ & 5\tabularnewline
				&  & $F_{2}^{(2;0,0)},\quad F_{3}^{(2;0,0)},\quad F_{4}^{(2;0,0)}$ &  & \tabularnewline
				&  & $F_{5}^{(2;0,0)}$ &  & \tabularnewline
				\cline{2-5} \cline{3-5} \cline{4-5} \cline{5-5} 
				& \{1;0,1\} & $F_{4}^{(1;0,1)}$ & $-$ & 1\tabularnewline
				\hline 
			\end{tabular}
		}
		\caption{Parity-violating GST monomials.}
		\label{tab:gstmono}
	\end{table}
\end{center}

For the $(1;1,0)$ category, there is one monomial
\begin{equation}
F_{1}^{(1;1,0)}=\frac{1}{\sigma^{3}}\varepsilon_{abcd}R_{ef}^{\phantom{ef} cd}\phi^{a}\phi^{e}\phi^{bf}.
\end{equation}
Here and in what follows, the factor $\sigma$ is a shorthand for $\sigma\equiv \sqrt{-\phi_{a}\phi^{a}}=\sqrt{2X}$. 

For the $(0;2,1)$ category, there is one monomial,
\begin{equation}
	F_{6}^{(0;2,1)}=\frac{1}{\sigma^{6}}\varepsilon_{abcd}\phi^{e}\phi_{f}\phi^{a}\phi_{e}^{b}\phi_{m}^{c}\nabla^{m}\phi^{df}.
\end{equation}
For the $(1;2,0)$ category, there are 10 linearly independent monomials,
\begin{align}
& F_{1}^{(1;2,0)}=\frac{1}{\sigma^{2}}\varepsilon_{abcd}R_{ef}^{\quad cd}\phi^{ae}\phi^{bf}, \label{app:F1201}\\
& F_{2}^{(1;2,0)}=\frac{1}{\sigma^{4}}\varepsilon_{abcd}R_{ef}^{\quad cd}\phi^{a}\phi^{e}\phi_{m}^{b}\phi^{fm},\\
 & F_{3}^{(1;2,0)}=\frac{1}{\sigma^{4}}\varepsilon_{abcd}R_{ef}^{\quad cd}\phi^{e}\phi^{m}\phi_{m}^{a}\phi^{fb}, \\ & F_{4}^{(1;2,0)}=\frac{1}{\sigma^{4}}\varepsilon_{abcd}R_{ef}^{\quad cd}\phi^{a}\phi^{m}\phi^{be}\phi_{m}^{f}, \\ & F_{5}^{(1;2,0)}=\frac{1}{\sigma^{4}}\varepsilon_{abcd}R_{ef}^{\quad cm}\phi^{a}\phi^{e}\phi^{bf}\phi_{m}^{d},\\ & F_{6}^{(1;2,0)}=\frac{1}{\sigma^{4}}\varepsilon_{abcd}R^{ae}\phi^{b}\phi^{f}\phi_{e}^{c}\phi_{f}^{d},\\
 & F_{7}^{(1;2,0)}=\frac{1}{\sigma^{6}}\varepsilon_{abcd}R_{ef}^{\quad cd}\phi^{m}\phi^{n}\phi^{e}\phi^{a}\phi_{m}^{f}\phi_{n}^{b},\quad \\
 & F_{8}^{(1;2,0)}=\frac{1}{\sigma^{6}}\varepsilon_{abcd}R_{ef}^{\quad cm}\phi^{a}\phi^{e}\phi_{m}\phi^{n}\phi_{n}^{b}\phi^{df},\\
 & F_{9}^{(1;2,0)}=F_{1}^{(1;1,0)}E_{1}^{(0;1,0)}=\frac{1}{\sigma^{4}}\varepsilon_{abcd}R_{ef}^{\quad cd}\phi^{a}\phi^{e}\phi^{bf}\phi_{m}^{m},\\
 & F_{10}^{(1;2,0)}=F_{1}^{(1;1,0)}E_{2}^{(0;1,0)}=\frac{1}{\sigma^{6}}\varepsilon_{abcd}R_{ef}^{\quad cd}\phi^{a}\phi^{e}\phi^{bf}\phi_{m}\phi_{n}\phi^{mn} , \label{app:F12010}
 \end{align}
where the 2 factorizable monomials $F_{9}^{(1;2,0)}$ and $F_{10}^{(1;2,0)}$ are also shown explicitly.
For the $(2;0,0)$ category, there are 5 linearly independent monomials,
\begin{align}
& F_{1}^{(2;0,0)}=\varepsilon_{abcd}R_{ef}^{\quad cd}R^{abef}, \label{app:F2001}\\
& F_{2}^{(2;0,0)}=\frac{1}{\sigma^{2}}\varepsilon_{abcd}R_{ef}^{\quad cd}R_{\quad\;m}^{abf}\phi^{e}\phi^{m},\\
 & F_{3}^{(2;0,0)}=\frac{1}{\sigma^{2}}\varepsilon_{abcd}R_{ef}^{\quad cd}R_{\quad\;m}^{efa}\phi^{b}\phi^{m},\\
 &F_{4}^{(2;0,0)}=\frac{1}{\sigma^{2}}\varepsilon_{abcd}R_{ef}^{\quad cd}R^{ae}\phi^{b}\phi^{f},\\
 & F_{5}^{(2;0,0)}=\frac{1}{\sigma^{4}}\varepsilon_{abcd}R_{ef}^{\quad cd}R^{amen}\phi^{b}\phi^{f}\phi_{m}\phi_{n}. \label{app:F2005}
\end{align}
For the $(1;0,1)$ category, there is a single independent monomial
	\begin{equation}
		F_{4}^{(1;0,1)}=\frac{1}{\sigma^{4}}\varepsilon_{abcd}R_{ef}^{\phantom{ef}cd}\phi^{a}\phi^{e}\phi^{m}\nabla^{b}\phi_{m}^{f}.
	\end{equation}

The above complete sets of GST monomials derived in \cite{Gao:2020juc} are independent in the sense of linear algebra. Some of the monomials are related to each other up to total derivatives.
For $F_{6}^{(0;2,1)}$, after some manipulations, we find
	\begin{equation}
		F_{6}^{(0;2,1)}=-\frac{1}{2}F_{7}^{(1;2,0)}+\nabla_{d}\left(\frac{1}{\sigma^{6}}\varepsilon_{abc}^{\phantom{abc}d}\phi^{e}\phi^{f}\phi^{a}\phi_{e}^{b}\phi^{cm}\phi_{mf}\right). \label{app:F0216red}
	\end{equation}
It is interesting to note that the total derivative has effectively vanishing contribution if the coefficients are functions of $\phi$ and $X$, since
\begin{eqnarray*}
	& & f\left(\phi,X\right)\nabla_{d}\left(\frac{1}{\sigma^{6}}\varepsilon_{abc}^{\phantom{abc}d}\phi^{e}\phi^{f}\phi^{a}\phi_{e}^{b}\phi^{cm}\phi_{mf}\right) \\
	& \simeq & \nabla_{d}f\frac{1}{\sigma^{6}}\varepsilon_{abc}^{\phantom{abc}d}\phi^{e}\phi^{f}\phi^{a}\phi_{e}^{b}\phi^{cm}\phi_{mf}\\
	& = & \left(\frac{\partial f}{\partial\phi}\phi_{d}+\frac{\partial f}{\partial X}\left(-\phi_{nd}\phi^{n}\right)\right)\left(\frac{1}{\sigma^{6}}\varepsilon_{abc}^{\phantom{abc}d}\phi^{e}\phi^{f}\phi^{a}\phi_{e}^{b}\phi^{cm}\phi_{mf}\right)\\
	& \equiv & 0.
\end{eqnarray*}
As a result, (\ref{app:F0216red}) implies that
\begin{equation}
	f\left(\phi,X\right)F_{6}^{(0;2,1)}\simeq-\frac{1}{2}f\left(\phi,X\right)F_{7}^{(1;2,0)}.
\end{equation}
For $F_{4}^{(1;0,1)}$, we have
	\begin{equation}
		F_{4}^{(1;0,1)}=-F_{2}^{(1;2,0)}-F_{4}^{(1;2,0)}-4F_{7}^{(1;2,0)}+\nabla^{b}\left(\frac{1}{\sigma^{4}}\varepsilon_{abcd}R_{ef}^{\phantom{ef}cd}\phi^{a}\phi^{e}\phi^{m}\phi_{m}^{f}\right).
	\end{equation}
For coefficients that are functions of $\phi$ and $X$, we have
	\begin{equation}
		fF_{4}^{(1;0,1)}\simeq-fF_{2}^{(1;2,0)}-fF_{4}^{(1;2,0)}-4\left(f-\frac{X}{2}\frac{\partial f}{\partial X}\right)F_{7}^{(1;2,0)}. \label{app:F1014red}
	\end{equation}
As a result, according to Table \ref{tab:gstmono}, since $F_{6}^{(0;2,1)}$ and $F_{4}^{(1;0,1)}$ can be reduced by integrations by parts, we are left with 15 GST monomials $F_{1}^{(1;2,0)},\cdots,F_{10}^{(1;2,0)}$ and $F_{1}^{(2;0,0)},\cdots, F_{5}^{(2;0,0)}$.
It is also interesting to verify that these 15 GST monomials satisfy 7 identities in the unitary gauge, which we choose to be:
\begin{eqnarray}
	F_{1}^{(2;0,0)}+4F_{3}^{(2;0,0)} & \xlongequal{\mathrm{u.g.}} & 0, \label{app:ugid1}\\
	F_{2}^{(2;0,0)}-4F_{5}^{(2;0,0)} & \xlongequal{\mathrm{u.g.}} & 0,\\
	F_{3}^{(1;2,0)}+F_{4}^{(1;2,0)}-2F_{6}^{(1;2,0)} & \xlongequal{\mathrm{u.g.}} & 0,\\
	F_{1}^{(1;2,0)}-2F_{4}^{(1;2,0)}-4F_{5}^{(1;2,0)} & \xlongequal{\mathrm{u.g.}} & 0,\\
	F_{3}^{(1;2,0)}-F_{7}^{(1;2,0)}+2F_{8}^{(1;2,0)}+F_{10}^{(1;2,0)} & \xlongequal{\mathrm{u.g.}} & 0,\\
	F_{1}^{(2;0,0)}-8F_{4}^{(2;0,0)}+8F_{1}^{(1;2,0)}-16F_{4}^{(1;2,0)}+32F_{6}^{(1;2,0)} & \xlongequal{\mathrm{u.g.}} & -16\mathcal{D},\\
	F_{2}^{(2;0,0)}+3F_{1}^{(1;2,0)}+2F_{2}^{(1;2,0)}+6F_{3}^{(1;2,0)}-2F_{4}^{(1;2,0)}-2F_{9}^{(1;2,0)} & \xlongequal{\mathrm{u.g.}} & -4\mathcal{D}, \label{app:ugid7}
\end{eqnarray}
where $\mathcal{D}$ in the last two identities is the total derivative term defined in (\ref{app:calD}).
Therefore, in the sense of the unitary gauge, there are only 8 independent GST monomials of $d=4$, which is exactly the same number of SCG monomials of the complete basis.

\providecommand{\href}[2]{#2}\begingroup\raggedright\endgroup

\end{document}